\documentclass[11pt]{article}
\usepackage{geometry}                
\usepackage{amssymb}
\usepackage{amsmath}
\usepackage{hyperref}
\usepackage{url}
\usepackage{color}

 \def\bleu {\color{blue}}
\newcommand{\blit}[1] {{\bleu \emph{  #1}}}

\def\gr {general relativity}

\def\Liegl {\underline{\mathfrak{g}\ell}}
\def\Lieso {\underline{\mathfrak{so}} }

\def\Fr {\mbox{Fr}}
\def\Sect {\mbox{Sect}}
\def\Minkv {Minkowski vector space}
\def\Minkvs {Minkowski vector space}
\def\LC {Levi-Civita}

\def\fb {fiber bundle}
\def\fibre {\mapsto}
\def\kg {Klein-Gordon}
\def\End {\mbox{End}}
\def\bbbZ{{\rm Z}}
\def\imply {\Rightarrow}
\def\implies {\Rightarrow}
\def\tensor#1#2#3{{#1^{#2}_{#3}}}
\def\bbbL{{\rm I\!L}}
\def\eqbydef {\stackrel{def}{=}}
\def\SL  {\mbox{SL}}
\def\Sp  {\mbox{Sp}}
\def\SUdd {\mbox{SU}(2,2)}

\def\I{{\rm I\!I}}
\def\mlr {Marc Lachi\`eze-Rey}
\def\spt {space-time}
\def\undemi {\tfrac{1}{2}~}
\def\ie {{i.e.}}
\def\eg {{e.g.}}
\def\Skew  {\mbox{Skew}}
\def\bydef{by definition}

\def\clif {\mbox{C}\ell }
\def\cliff {\mbox{C}\ell }
\def\setC{\mathbb{C}}
\def\cclif {\setC  \ell }
\def\ccliff {\setC \ell }
\def\setH{{\rm I\!H}} 
\def\Id{\mathbb{I}}
\def\Cliff {\mbox{C}\ell }
\def\R{{\rm I\!R}}
\def\andy{\mbox{~and~}}
\def\calI {{\cal I}}
\def\guill {\textquotedblleft ~}
\def\setZ{\rm Z}  
\def\hodge { {}^{\star}}
\def\Hodge { {}^{\star}}
\def\with  {~\mbox{with}~}
\def\EQUIV {\Leftrightarrow}
\def\diese {\sharp}

\def\Vol {\mbox{Vol}}
\def\calM {{\cal M}}
\def\T{{\rm T}}
\def\bbbM{{\rm I\!M}} 
\def\Mk{{\rm I\!M}} 
\def\bbbMc{{\rm I\!M _{\setC} }} 
\def\Mkc{{\rm I\!M _{\setC} }} 
\def\mink {Minkowski}
\def\Mink {Minkowski}
\def\minks {Minkowski spacetime}
\def\Minks {Minkowski spacetime}
\def\Mat {\mbox{Mat}}
\def\coord {coordinate}
\def\myspan {\mbox{span}}
\def\Herm {\mbox{Herm}}

\def\trace {\mbox{Tr}}
\def\tr {\mbox{Tr}}
\def\calO {{\cal O}}
\def\SO  {\mbox{SO}}
\def\SOf {\mbox{SO}^{\uparrow}}
\def\Pin {\mbox{Pin}}
\def\Spin {\mbox{Spin}}
\def\Spinf {\mbox{Spin}^{\uparrow}}
\def\uc  {universal covering}
\def\d {{\rm d}}

\def\Fr {\mbox{Fr}}
\def\FrGL {\mbox{Fr}^{\mbox{{\scriptsize    GL}}}}

\def\FrSO {\mbox{Fr}^{\mbox{{\scriptsize     SO}}}}
\def\FrSOf {\mbox{Fr}^{\mbox{{\scriptsize     SO}}\uparrow}}

\def\spb {  {\mathsf{o}}}
\def\fibre {\mapsto}

\def\OA {\calO  ^{A}}
\def\OAdot {\calO  ^{\dot{A}}}
\def\OAAdot {\calO  ^{A\dot{A}}}

\def\Adot {\dot{A}}
\def\Bdot {\dot{B}}
\def\dotA {\dot{A}}
\def\dotB {\dot{B}}
\def\undot {\dot{1}}
\def\deuxdot  {\dot{2}}

\def\undemi {\tfrac{1}{2}~}
\def\unquart {\tfrac{1}{4}~}

\newcommand{\binomial}[2]{\left(\begin{array}{c}#1\\#2  \end{array} \right)}
\newcommand{\inner}[2]{\langle #1 \mid #2 \rangle}
\newcommand{\vectord}[2]
{\begin{pmatrix}
#1   \\
#2 
\end{pmatrix}}
\newcommand{\matrixdd}[4]
{\begin{pmatrix}
#1    &  #2 \\
#3    &  #4
\end{pmatrix}}
\newcommand{\pder}[2]{\frac{\partial#1}{\partial#2}}

\title{Spin and Clifford algebras, an introduction}
\author{\mlr\\
Laboratoire
\emph{AstroParticules et Cosmologie}\\ UMR 7164 (CNRS, Universit\'e Paris 7, CEA, Observatoire de Paris)
}

\begin{document}
\maketitle
{\bf Abstract} In this  short pedagogical presentation, we introduce the    spin groups and the   spinors from the point of view of group theory. We also present, independently, the construction of the low dimensional Clifford algebras. And we establish the link between the two approaches. Finally, we give some notions of the generalisations to arbitrary \spt s, by the introduction of the spin and spinor bundles. 

\tableofcontents

 
\section{Clifford algebras}

A    Clifford algebra is canonically associated to any   vector space $(V,g)$ with a quadratic form $g$ (a scalar product). This  algebra,  compatible with the quadratic form,   extends the capacities of calculations on $V$.
One distinguishes real and complex  Clifford algebras, which extend real and complex 
 vector spaces. 

\subsection{Preliminaries: tensor algebra and exterior  algebra over a vector space}\label{Preliminaries}

\subsubsection{Tensor algebra  over a vector space}\label{tensor}

Let us consider    a vector space $ V$ (no scalar product is assumed).
From $ V$ and its dual $ V ^*$ (space  of one-forms), are   constructed  new objects called tensors, which form an algebra. Among the tensors, we will  select  the completely antisymmetric ones,   called \blit{multivectors} or  \blit{multiforms}.  Multivectors form the exterior   algebra of  $ V  $. Multiforms  form the exterior   algebra  of the dual vector space  $ V ^*$.

 We recall the definition of the 
the vector space  of tensors  of type $(s,r)$:
 $$\tau^{(s,r)} V\equiv    \bigotimes ^s V^* \bigotimes ^r V,$$  with $s$   and $r$ factors respectively. 
Vectors are  (0,1) tensors; one-forms are (1,0) tensors. 
An $(s,r)$ tensor is a  linear operator  on    $V^s \otimes (V^*)^r$, and this may be  taken as the definition.

A basis (frame) $(e_A) $  for $V$ induces canonically a  reciprocal basis (coframe) $(e^A) $  for $V^*$. Their tensor products  provide   a canonical  basis $(e_{A_1}) \otimes (e_{A_2}) \otimes ... (e_{A_s}) \otimes  (e{^B_1}) \otimes (e^{B_2}) \otimes ... (e^{B_r})$   for tensors. 
This extends the   isomorphism between $V$ and $V^*$ to an   isomorphisms between $ \tau ^{(r,s)}$ and $ \tau ^{(s,r)}$.

By the direct sum operation, one defines  the vector space of all tensors $$\tau V = \bigoplus  _{s=0,r=0}^\infty \tau^{(s,r)}V.$$ It  has a (non commutative)  algebra structure with respect to the tensor  product. 

The sets of all tensors of $(s,0)$ type, $\forall s$ (called covariant), and    
of all tensors of $(0,r)$ type, $\forall r$ (called contravariant) have similar vector space structures.
These definitions  of  tensors  require 
 no other structure than that of the vector space.

Now we will define   antisymmetrisation properties of tensors:
An antisymmetric [contravariant]Êtensor of type $(0;p)$  will be called  a   \blit{$p$-vector}, more generally a   \blit{multivector}.
An antisymmetric  [covariant]Ê tensor of type $(p;0)$  defines a 
 \blit{$p$-form}, more generally a   \blit{multiform} (more simply, a form).

\subsubsection{Antisymmetry and the wedge  product}\label{bivector}\label{multivector}\label{wedge}

Given  a  vector space   $V$, the (normalized)  antisymmetric part of  the tensor product of two vectors is defined as 
$$v \wedge w=\undemi ~Ê(v \otimes w-w \otimes v).$$
(care must be taken that one often finds the definition without the normalizing factor).
This is an antisymmetric tensor of rank (0,2), also called a  \blit{bivector}. The goal of this section is to extend this definition, \ie, to   define the antisymmetric part of  a  tensor product of an arbitrary number of  vectors. This  defines     a new product,   the  \blit{wedge product}.  The wedge product of two  vectors  defines a bivector. Its generalization will lead  to consider    new objects
  called \emph{multivectors} (= skew contravariant tensors).

The wedge product is also defined for the dual $V^*$. In the same way that the vectors of $V^*$ are called the one-forms of $V$, 
 the multivectors of   $V^*$ are called the     \emph{multi-forms} (= skew covariant tensors) of $V$, which are usually be simply called \emph{forms}.   
 
    With the wedge product, multivectors form an algebra, the  \blit{exterior algebra} [of multivectors]  $\bigwedge V$of $V$.  Multiforms form the exterior algebra of multiforms  $\bigwedge V^*$ on $V$. 
    These  algebras are defined in the absence of any inner product or metric in the initial vector space. However, an inner product will allow us  to define   additional structures:
 \begin{itemize}
  \item a canonical (musical) isomorphism between  $V$and its dual $V^*$, which extends to the   exterior algebras $\bigwedge V$  and $\bigwedge V^*$;
  \item an (Hodge) duality (\ref{Hodge}) in the exterior algebras;
  \item an  additional  algebra structure:  that of Clifford algebra, which   result  from the   definition of  new products  on  $\bigwedge V$  and $\bigwedge V^*$, the Clifford  products.
\end{itemize}   

{\bf The antisymmetric symbol}\label{antisymmetric symbol}


To define properly the wedge product, we    introduce  the antisymmetric symbol.
Let us consider the set $  \{1,2,...,n\}$  of the $n$ first integers. We recall that a \emph{permutation} is an ordered version of this set,  $(i_1,i_2,...,i_n)$, where each $i_k \in    \{1,2,...,n\}$.
Its parity is defined as the number of pair exchanges necessary to reach it  from the permutation 
$1,2,...n$. 

We define    the completely antisymmetric symbol $[i_1,i_2,...,i_n]$,   which takes the value 1,-1, or 0, according to the parity of the permutation $(i_1,i_2,...,i_n)$.  We have for instance $$[1,2,3,..., n]=1,~[2,1,3,..., n]=-1,~[1,1,3,..., n]=0.$$

Note that  the number of non zero permutation is $n!$. One often   writes       $ [\alpha, \beta, \gamma, ...]$ under the form    $ {\epsilon}_{\alpha \beta  \gamma  ...}$.  

\subsubsection{The  operator of antisymmetry and the wedge product}

If $V$ is a vector space of dimension $d$, the tensor product $$ \bigotimes ^p V= V \otimes V \otimes ...\otimes V $$ has also  a vector space structure. Its   elements, the tensors of type (0,p), are sums of elements  of the form  $v_1 \otimes v_2\otimes ... \otimes v_p$.

To such a tensor, we associate its (normalized) completely  antisymmetric part
\begin{equation}
\label{skew}
v_1 \wedge v_2 \wedge ... \wedge v_p \equiv
 \Skew[v_1 \otimes v_2\otimes ... \otimes v_p] \equiv
  \frac{ \sum  [i_1,i_2,...,i_p]  ~v_{i_1} \otimes v_{i_2}  \otimes ...  \otimes    v_{i_p}}{p!}.\end{equation}
The  sum extends   over all permutations (we recall that  $p! = \sum  [i_1,i_2,...,i_p]$).
It is called the  \blit{wedge (or external) product}.    


     Such an external product  is a   skew $(0,p)$ tensor  called a $p-$multivector (or $p-$vector). The definition is extended by linearity : the sum of two $p-$multivectors is a $p-$multivector:  the $p-$multivectors form  the vector space  $   \bigwedge ^p V $,   of  dimension  $\binomial{d}{p}$ (the binomial coefficient).

If $V_p$ and $V_q$ are a $p-$vector and a $q-$vector, we have
$$ V_p \wedge  V_q = (-1)^{pq}~ V_q \wedge  V_p.$$

[Note that the wedge product is often defined without normalization.  In this case, many formulas  differ by the     factor $p!$].

For instance, the external product of two vectors is   the antisymmetrical part of their tensor product:
$$ v \wedge w \equiv  \frac{   {v \otimes w- w \otimes v}}{2}.$$  It results that $v \wedge v =0$.  
    
The  wedge product of vectors  is  distributive, associative and completely antisymmetric.
The wedge product of a   number $p$  of vectors  is zero iff the vectors are linearly  dependent. 
This implies that the maximum order of a multivector is $d$, the dimension of the original vector space,  and also that there is only one multivector of order $d$, up to a multiplicative constant.

A scalar is identified as a multivector of order zero.
An usual vector  is  a multivector of order one.
The wedge product of a $p-$mutivector by itself, $M \wedge M$, is always 0 when $p$ is odd. This is not true when $p$ is even.

We will  give to multivectors     an algebraic structure   by extending  the external   product to  them.

\subsubsection{The exterior algebra of multivectors}

The bivectors form the vector space $\wedge  ^2(M)$, of dimension $\frac{d(d-1)}{2}$.  
 A general bivector cannot necessarily be decomposed as a wedge product. When this is  possible, the bivector is called  \blit{ simple}.
A simple bivector $B=a\wedge b$  can be considered as the oriented triangle with the  vectors $a$ and  
$b$  as sides. Then $ \inner{B}{B}$ (see below) is the oriented area of the triangle.

Now   we extend the sum to  multivectors of different orders, up to $d$, like $$A_0 +A_1 +A_2+...+A_d,$$ where $A_p$ is a $p$-vector (the expansion stops  at $d$).  Such multivectors have not a definite order. They belong to the vector space $$\bigwedge V \equiv \bigoplus _{p=0}^d \bigwedge ^p V$$ of all  multivectors, of   dimension $2^d$.

The wedge product is easily extended to all multivectors by linearity, associativity,  distributivity and anticommutativity for  the  1-vectors. This provides an an algebra structure to
$\bigwedge  V$: the  \blit{exterior algebra of multivectors}.  

This also  
   allows the practical calculations of wedge products. For instance, $$
v \wedge w \wedge (v+w+x)=
v \wedge w \wedge v+
v \wedge w \wedge  w+
v \wedge w \wedge  x $$ $$=
-(v \wedge v )\wedge w+
v \wedge( w \wedge  w)+
v \wedge w \wedge  x =
v \wedge w \wedge  x ,$$ which is a trivector if we assume $v$, $w$ and $x$ linearly independent.

A   scalar  product on  $V$ allows us to define an other  algebra structure  for the  multivectors. This results from the introduction of  new  product, the Clifford product, which unifies the wedge product and the scalar product : the Clifford  algebra of multivectors,  presented in (\ref{Clifford algebra}). 
Contrarily to the wedge product, the Clifford product is, in some cases,  invertible.

\subsection{Clifford algebras}\label{Clifford algebra}

\subsubsection{The Clifford product}

The definitions of the wedge product, and of the  multivectors do not depend on any inner product.  
Now, we will   assume an   inner   product    in $V$:  $$g:~Êu,v \mapsto  g(u,v) \equiv u  \cdot v  .$$
 
 One defines  the  \blit{Clifford   (or   {geometrical})   product} of two vectors as $$u~v  \equiv u \cdot v +u \wedge v.$$ In general, this appears as the sum of a scalar   (polyvector of grade zero)   
plus  a bivector (polyvector of grade 2), thus a \emph{non homogeneous}   multivector. The scalar product $u \cdot v=\frac{u~v+v~u}{2}$
 and the wedge product 
    $u \wedge v=\frac{u~v-v~u}{2}$ appear as the    symmetrical  and
  antisymmetrical parts of the Clifford product. 

 The Clifford product is \bydef ~associative and distributive. 
These properties allow us  to  extend it  to all multivectors.  
To illustrate:
$$ a b a =a (ba)=a~(b\cdot a+b\wedge a)=
a~(a\cdot b-a\wedge b)=
a~(2a\cdot b-a  b)=
2a~( a\cdot b) -a a  b =
2 ~( a\cdot b)~a -(a\cdot a)  b .$$

or  $$
v ~ w ~ (v+w+x)=
v ~ w~ v+v ~ w~  w+v ~ w~  x =
2 ~( v\cdot w)~v -(v\cdot v)  w
+v ~ (w~  w)+
v ~ w~  x $$ $$=
v ~( w\cdot v+w\wedge v)+
v ~ (w\cdot w+w\wedge w)+
v ~ w~  x $$ $$=
v ~( v\cdot w-v\wedge w)+
v ~ (w\cdot w)+
v ~ w~  x =
v ~( v ~ w)+
v ~ (w\cdot w)+
v ~ w~  x $$ $$=
(v ~v) ~ w)+
v ~ (w\cdot w)+
v ~ w~  x =
 (v\cdot v ) ~ w +
v ~ (w\cdot w)+
v ~ w~  x .$$
This polyvector is a sum of vectors (polyvector of grade 1) and trivectors  (polyvector of grade 3).
  
\subsubsection{The Clifford algebra}

The     \blit{Clifford algebra} $\clif  (V)$ is defined as $\bigwedge V$, with the   Clifford product    $ v,w \mapsto   v  ~Êw$. 
As a vector space (but not as an algebra), $\clif  (V)$ is isomorphic to the exterior algebra $\bigwedge V$. Thus, its elements are  the multivectors defined over $V$, although with a different multiplication law which takes into account the properties of the metric.  It  provides    an extension of $V$, and  of the   the calculation possibilities.   (In the special case where  the metric is zero,  $\clif  (V)=\bigwedge V$.)

More formally, one may define  
$\cliff(V)$ as the quotient of the tensor algebra $T(V)$ over $V$ by the  (bilateral) ideal   generated by the set $\{x \in V; ~x \otimes x-g(x,x)~Ê\Id \}$.

Note that the Clifford algebra structure may be defined in an abstract way, with a Clifford product. 
In this case, the vector space of multivectors is simply a peculiar representation. Here we will present the Clifford algebra structure through this representation. Other representations also exist.  

A polyvector of definite order is called homogeneous. In general, this is not the case, and we define the projectors $<\cdot>_r$ which project a polyvector onto its homogeneous part of grade $r$.

We call  $\Cliff ^k(V)$  the vector space of polyvectors of grade $k$. 
  As a vector space, we have $$\Cliff(V)=\bigoplus _{k=0}^d \Cliff ^k(V).$$

As vector spaces, we have $\Cliff ^0(V) \equiv \R$, which is thus   seen as embedded in   $\clif  (V)$, as the   multivectors of grade 0 (0-vectors). 
The vector space $V$  itself may be seen as embedded in   $\clif  (V)$, as 
 $\Cliff ^1(V) $: its elements are the multivectors of grade 1 (or 1-vectors).

 {\bf Paravectors}

The addition of a scalar plus a grade one vector is called a  \blit{paravector}. It can be expanded as  $A=A^0 + A^i~e_i$, where $A^0=<A>_0$ and $A^i~e_i=<A>_1$.
The  vector space of paravectors is thus $\R \oplus V=  \clif^0(V)\oplus  \clif^1(V)\subset \clif(V)$.
 
 We define also the even and odd  subspaces of a Clifford algebra  $C$ as  the direct sums $$C^{even} =\bigoplus _{k ~even}  C ^k  \andy C ^{odd} =\bigoplus _{k ~odd} C ^k .$$ Both 
have dimension $2^{d-1}$ and  $C^{even}$ is a subalgebra of $C$. 

{\bf The pseudoscalars}\label{}
 
 Up to a multiplicative scalar,  there is a unique $d$-multivector. To normalise, we   choose  an oriented ON basis for $V$,  and define   $\calI= e_{1}\ldots  e_{d} = e_{1}\wedge \ldots \wedge   e_{d} $ as  the  \blit{orientation operator}. 
It  verifies  $$\calI^{2}=(-1)^{\frac{d(d-1)}{2}+s},$$  depending on the dimension and on the signature of the vector space $(V,g)$.  
The multiples of $\calI$ are   called the  \blit{pseudoscalars}. 

When the dimension is odd, $\calI$ commutes with all multivectors. When the dimension is even, it commutes with even grade  multivectors, and anti-commutes with odd grade ones: 
$$ \calI ~P_r=(-1)^{r(d-1)}~P_r ~\calI.$$
The  \blit{center} (the set of elements commuting with all 
elements)   of $\cliff(V)$  is $\cliff^0(V)$ for $d$ even, 
 or $\cliff ^0(V) \oplus \cliff^d(V)$  for $d$ odd.

The multiplication rules imply that   the multiplication by $\calI$ 
transforms  a grade $r$ polyvector $P_r$ into the  grade $d-r$ polyvector $\calI P_r$,  called the   \blit{orthogonal complement} of $P_r$.

{\bf Bivectors}
                         
After the scalars and the 1-vectors, the bivectors  are the  simplest polyvectors.     
The wedge product of two bivectors  is zero or a quadrivector.  

A bivector is called \guill simple " (or  decomposable) if it can be written as a wedge product. 
Not all bivectors are simple, and one defines the rank of a bivector $B$ in the following equivalent ways:
\begin{itemize}
  \item 
i) The minimum integer $r$ such that $\wedge ^rB=0$.
  \item 
ii) The minimum number of non-zero vectors whose exterior products can add up to form $B$.
  \item iii) The number $r$ such that the space $\{ X\in V;  X \wedge B = 0\}$ has dimension~$r$.
\item iv) The rank of the component matrix of $B$, in  any frame on $V$.  \end{itemize}


A bivector $B$ of  rank    $2$ (minimum value for a non zero bivector)  is simple. 
The  \blit{simplicity condition} is expressed as $B \wedge B=0$ or, in tensorial components,  \begin{equation}
\label{ }
B_{[\mu\nu }~B_{\rho\sigma] }=0.Ê
\end{equation}
For instance,  the Plebanski formulation of \gr ~\cite{Plebanski}Ê considers a bivector  as the dynamical variable, to which is imposed a simplicity constraint.

To each simple bivector, one may associate uniquely, up to a scalar multiplication,
  the [two-]plane span$(v,w)$ through the origin, subtended by two vectors $v$ and $w$.  
  This establishes  a one to one correspendence  between the  planes through the origin and    the projective simple     bivectors $[v\wedge w]$, where $[v\wedge w]$ is defined as the set of bivectors  proportional to the bivector  $v \wedge w$. The projective simple     bivector  $[A \wedge B]$ belongs  to the projective space $P\bigwedge ^2(V)$, the set of equivalence classes of bivectors under the scalar multiplication. 
              
 \subsubsection{Automorphisms in Clifford algebras}

 There are three important automorphisms canonically defined on a Clifford algebra $C$.
 \begin{itemize}
  \item  {\bf The reversion}\label{reversion}:\\
the \blit{reversion}, or \emph{principal anti-automorphism} is defined as the   transformation $C \mapsto C$ which reverses the 
order of the factors in any polyvector:
 $$R \equiv v_{1} \ldots  v_{k} \mapsto  {R}^T \equiv v_{k} \ldots  v_{1}.$$  It is trivially extended by linearity. Scalars  and vectors remain unchanged.  Bivectors change their sign. For an homogeneous multivector, we have 
 $$(A_r)^T=(-1)^{r(r-1)/2}~ÊA_r,~Êr\ge 1.$$
  \item  {\bf Main involution}: \\
  the \blit{main involution} (or grade {involution}) $a \mapsto a^*$ is defined through its action  $ e_i \mapsto -e_i$  on the vectors of $\Cliff^1$. It may also be written $$a \mapsto (-1)^d~Ê\calI~a~\calI^{-1}.$$ Even  or odd grade elements of $C$ form the two eigenspaces $C^{even}$ and $C^{odd}$, with eigenvalues 1 and -1, of the grade involution.
  Noting  that 
  $$C =  C^{even}\oplus  C^{odd},$$
$$ C^{even}~Ê C^{even} =C^{odd}~Ê C^{odd} =C^{even};
~Ê  C^{even}~Ê C^{odd} =
   C^{odd}~Ê C^{even} =C^{odd}$$
makes $C$ a $\setZ _2-$graded algebra.   
   
  \item  {\bf [Clifford] conjugation}: \\ 
  the  \blit{conjugation}, or antiautomorphism, is the composition of both:
$$  \bar{R}= ({R}^*)^T.$$
  \end{itemize}

\subsubsection{Scalar product of multivectors and Hodge duality}\label{Hodge}
 
The scalar product of $V$ is extended to $\Cliff(V)$  as
$$g(A,B) = A  \cdot B =<A^T ~ B >_0,$$
where $<\cdot>_0$ denotes the scalar  part.

 It is bilinear. It reduces to zero  for  homogeneous multivectors of different grades.  
It reduces to the usual product  for scalars (grade 0), and to the metric product for 1-vectors   (grade 1).
In general, we have the decomposition
$$A\cdot B = <A>_0 \cdot <B>_0 + <A> _1 \cdot <B>_1 +...+ <A>_n \cdot <B>_n.$$

{\bf The  Hodge duality}\label{HodgePolyvectors}

The \blit{Hodge duality}  is defined as the operator \begin{eqnarray}
\hodge : & \wedge ^p &\mapsto \wedge ^{n-p} \\
  & A_p &\mapsto \hodge A_p 
\end{eqnarray}such that
$$B_p \wedge ( \hodge A_p) =( B_p \cdot A_p)~\calI,~\forall   B_p \in \wedge ^p .$$
It may be checked that, for $p$-forms, it coincides with the usual Hodge duality of forms defined from  the metric.

The simplicity condition  for a bivector can be written as
 $B \wedge B=0 \EQUIV <B , \Hodge B>=0$, implying  that $ \Hodge B$ is also simple. 

In 4 dimensons, the  Hodge duality  transforms a bivector into a bivector.
Any bivector can be decomposed in a  \blit{self-dual} and an \blit{anti-self-dual} part:
$$B = B^{+} + B^{-},~\Hodge{B} = B^{+} - B^{-}.$$

\subsubsection{Frames}

A frame $(e_i)_{i=1 \cdot \cdot \cdot n}$ for $V$ defines a natural  frame for $\bigwedge  V$. To define it, we consider all the finite sets of the form 
$$I \equiv \{i_1,  ..., i_k \} \subset  \{1,  ..., n \}, \with i_1 < i_2 <...<i_k.$$
We define the  multivectors $e_I=e_{i_1} \wedge  ...
 \wedge  e_{i_k}$, and $e_{\emptyset}=e_0 =1$ (ordered sequences only).
 The   multivectors $e_I$  provide a   basis for the vector space $\bigwedge V$, and thus for $C$, with  the
  \guill orthographic " index $I$ going from 1 to $2^n$.
   
A multivector is expanded in this basis as
 $$A =A^I~e_IÊ\equiv A_0+A^i~Êe_i+A^{ij}~Êe_{\{ij\}}+...+A^{1,2,...,n}~Êe_{\{1,2,...,n\}}.$$ 
Its  components $A^{I}$   may be seen as  coordinates in $C$. Thus, functions on $C$ may be considered as  functions of the coordinates, and this allows us to  define a  differential structure, with a basis for one-forms given by the 
$\d X^{A}$.

 When  the  basis $(e_i)$ is  ON ($e_i \cdot e_j =\eta_{ij}=\pm \delta_{ij}$), it is so for the   basis $(e_I)$  of $\Cliff(V)$, and we may   define    extended metric coefficients
 $\eta _{IJ} \equiv e_I \cdot e_J .$ 
 In such an ON basis, the scalar product of arbitrary multivectors expands as  $$A \cdot B = 
  \eta_{IJ}~A^I~B^J \equiv  A^0~ÊB^0\pm A^i~B^i\pm   A^{ij}~B^{ij}\pm ...\pm   A^{1,2,...,n}~B^{1,2,...,n}.$$ Summation is assumed over all orthographic indices, and   the $\pm$ signs depend on the signature.  

\subsection{Vectors and forms}

Given a vector space $V$,  we recall that its  \blit{dual} $V^*$ (the set of linear forms on $V$) is a vector space isomorphic to $V$.
A Clifford algebra may be similarly constructed from  $V^*$. Thus, to $(V,g)$ one associates its  Clifford algebra of multivectors $\clif(V)$, and its   Clifford algebra of [multi-]forms $\clif(V^*)$. They are isomorphic. A bivector of 
$\clif(V^*)$ is called a 2-form of $ V $, etc.
 
The scalar product induces the  \blit{canonical (or musical)  isomorphism} 
between $V$ and $V^*$.
It  is easily extended to  
 an  isomorphism between $\clif (V)$ and $\clif (V^*)$. 
The   scalar product $g$ of $V$ induces the scalar product on  $V^*$ (also written $g$) $$g(\alpha,\alpha)=g(\diese\alpha,\diese\alpha).$$ It is extended to $\clif (V^*)$ as above.
The pseudoscalar  $\calI$ of $\clif (V^*)$ identifies with the volume form $\Vol$ associated to the metric. 
The   Hodge duality (\ref{Hodge}) in  $\clif (V^*)$  identify  with its  usual definition    for forms.

In a [pseudo-]Riemannian manifold $\calM$,  the tangent spaces $\T _m \calM$, and their duals  $\T ^*_m \calM$, at all   points $m \in \calM$ define the tangent and the cotangent  bundles. Similarly, the reunions of their Clifford algebras define the  \blit{Clifford bundles} of multivectors and multiforms on $\calM$, respectively (see below).

 \subsection{Complex Clifford algebra}

When the vector space  is a complex vector space, its Clifford algebra is also complex. 

Given a real vector space $V$, we note $\cclif (V) $ the complexified Clifford algebra $\setC \otimes \clif (V) $. A case of interest for physics is when $V=\R^{1,3}=\bbbM$, the  \Minkv, and we  study below the    \blit{\spt -algebra}   $\cliff (\bbbM) $. 
Its complexification   $\cclif (\bbbM) \equiv \cclif(\R^{1,3} )$ is called the  \blit{Dirac algebra}. It  is isometric to $\clif(\R^{2,3})$, and $\clif(\R^{1,3})$ appears as a   subalgebra of $\clif(\R^{2,3})$.

More generally, from the complex algebra $\cclif(n) $ it is possible to extract the real Clifford algebra
$\clif(p,q) $ with $p+q=n$. To do so, we extract $\R^{p,q}$ from $\setC^n$:  as a complex space, $\setC$ admits the basis  $e_1,..., e_n$. We may see $\setC$ as a real vector space  with  the basis 
$e_1,..., e_p,i e_{p+1},..., i e_{p+q}$. Chosing $n$ vectors in this list, we construct the real subvector space   $ \R^{p,q}$, which heritates from the quadratic form. 
It follows that any element $a \in \cclif(n) $ may be decomposed as $a=a_r+i a_c,~Êa_r, a_c \in \clif(p,q)$ (see more details in, \eg,  \cite{Varlamov}).

{\bf Matrix representations} 
 
There are natural representations of $\cclif(d)$ 
on a (complex) vector space of dimension  $2^{k}$, with $k\equiv [d/2]$ (integer part)  \cite{Rausch}. Its elements   are called Dirac spinors, see below.
Elements of $\cclif(d)$ are represented  by matrices of order $2^{ k}$, \ie, elements of  the algebra  $\Mat _{2^k}( \setC)$,    acting as endomorphisms.

This representation is faithful when $d$ is even
and non-faithful when $d$ is odd. 
 
\subsection{The simplest Clifford  algebras}

The structure of a real   Clifford algebra is determined by the dimension of the vector space and the signature of the metric, so that it is written $\clif_{p,q}(\R)$. 
It is expressed  by its multiplication table. 
A  \blit{matrix representation}
of a  Clifford algebra is  an isomorphic algebra of matrices,  which thus  obeys the  same multiplication table. (Such matrix representations 
lead to the construction of spinors, see below).

The   table (\ref{first Clifford algebras})  gives   the matrix representations of   the lower dimensional     Clifford algebras. It is extracted from    
\cite{Rausch}, who gives its extension up to $d=8$.
Note the links with complex numbers and quaternions.

Periodicity theorems allow to explore the  Clifford algebras beyond   dimension 8. They  obey the following algebra  isomorphisms\begin{equation}
\label{Periodicity}
\cliff (p+1,q+1) \approx \cliff (1,1) \otimes \cliff (p,q),\end{equation}
$$\cliff (p+2,q) \approx \cliff (2,0) \otimes \cliff (p,q),$$ 
$$\cliff (p ,q+2) \approx \cliff (0,2) \otimes \cliff (p,q).$$

We will pay special attention to 
 \begin{itemize} 
\item the algebra of the  plane      $\clif  (\R ^{2}) =\clif  ( 2 ) $;
\item    the space algebra, or {\sl  Pauli algebra} $\clif  (\R ^{3})=\clif  (3)  $;
\item The \spt ~algebra $\clif  (1,3) $,
  the algebra  of [\mink] \spt, that we   describe    below in (\ref{sptalgebra}). 
Note the difference between $\clif  (1,3) $ and $\clif  (3,1) $ which may indicate a non complete equivalence  between the two signatures for \minks.
\end{itemize}

\begin{table}
\begin{tabular}{|c|c|}
	\hline
$\cliff(0,0)$&$\R$\\
	\hline
$\cliff(1,0)$&$\R \oplus \R$\\
$\cliff(0,1)$&$\setC$\\
\hline
$\cliff(2,0)$&$\Mat_2( \R)$\\
$\cliff(1,1)$&$\Mat_2( \R)$\\
$\cliff(0,2)$&$\setH$ (quaternions)\\
\hline
$\cliff(3,0)$&$\Mat_2(\setC)$\\
$\cliff(2,1)$&$\Mat_2(\R)\oplus M_2(\R)$\\
$\cliff(1,2)$&$\Mat_2(\setC)$\\
$\cliff(0,3)$&$\setH \oplus \setH$\\
\hline
$\cliff(4,0)$&$\Mat_2(\setH)$\\
$\cliff(3,1)$&$\Mat_4(\R)$\\
$\cliff(2,2)$&$\Mat_4(\R)$\\
$\cliff(1,3)$&$\Mat_2(\setH)$\\
$\cliff(0,4)$&$\Mat_2(\setH)$\\
\hline
\hline
  	\hline
\end{tabular}
	\caption{The first low dimensional  Clifford algebras (from  \cite{Rausch}).  $\Mat_n(K)$ denotes the algebra of $n\times n $ matrices with elements in $K$.}
\label{first Clifford algebras}
\end{table} 

\subsection{The geometric algebra of the plane} 

The   Clifford  algebra  of the plane, $\clif (\R ^{2})\equiv \clif ( 2 ) $ extends the  two-dimensional plane  $(\R^2,g)$, with the   Euclidean scalar product $g(u,v)=u \cdot v$.
Let us  use an ON basis ($e_{i}  \cdot e_{j} =\delta_{ij}$) for $\R ^{2}$. 

Antisymmetry implies 
that the only  bivector (up to a scalar) is $$e_{1} ~e_{2}=e_{1} \wedge  e_{2}=-e_2~Êe_1  \equiv \calI_{\clif ( 2 ) }=\calI.$$ The rules above imply $\calI ^{2}=-1$. We may check that $\clif(2)$  is closed for 
multiplication, and  admits the basis (1, $e_{1}$, $e_{2}, \calI)$,  as indicated in the table (\ref{ttttttttt}).

The general   polyvector expands   as
\begin{equation}
A=A^{0}~1+A^{1}~ e_{1}+A^{2}~ e_{2}+A^{3}~\calI,
\end{equation}  so that the four numbers $A^i \in \R$ play the role of \coord s for $\clif ( 2 ) $.
The full multiplication rules (see table \ref{The multiplication table for the algebra of the plane})  follow from associativity, symmetry and 
antisymmetry of the different parts. 

\begin{table}
	\begin{tabular}{|c|c|c|}
	\hline
	$j_{0} \equiv 1$ & $j_{1} \equiv  e_{1}$, $j_{2} \equiv  e_{2}$ &$j_{3} \equiv \calI \equiv e_{1}   e_{2} $ \\ \hline
	one scalar  & 2 vectors & one bivector   \\\hline
\end{tabular}
	\caption{The basis of the algebra of the plane}
\label{ttttttttt}
\end{table} 

\begin{table}
 \begin{tabular}{|c|c|c|c|}
	\hline
	1 & $e_{1}$& $e_{2}$ &$\calI $ \\
	\hline
	$e_{1}$& 1& $\calI $&   $e_{2}$ \\
	\hline
	$e_{2}$ &- $\calI$ &1&$-e_{1}$ \\
	\hline
	$\calI$  & $-e_{2}$& $e_{1}$ &$-1  $ \\
	\hline 
\end{tabular}
	\caption{The multiplication table for the algebra of the plane}
\label{The multiplication table for the algebra of the plane}
\end{table} 

{\bf The algebra of the plane and complex numbers}

The Euclidean plane  $\R ^2$ is naturally embedded (as a vector space)  in 
$\clif  (2) $ as $\clif ^1(2) $, the set of 1-vectors. 
We have the embedding  isomorphism 
\begin{eqnarray}
  & \R ^2   & \mapsto \clif ^1(2)  \\
  & (x,y) & \mapsto x ~e_{1}+y~e_{2}.
\end{eqnarray}
Since  $\R ^2$ may be seen as $\setC$, this  may also be written 
\begin{eqnarray}
  & \setC   & \mapsto \clif ^1(2)  \\
  &  x+i~y  & \mapsto x ~e_{1}+y~e_{2}.
\end{eqnarray}
    The right   multiplication of  such  a 1-vector by $\calI $ gives another 1-vector: $(x~e_{1}+y~Êe_{2})~\calI =x~e_{2}-y~e_{1}$.  We recognize      a rotation by $\pi/2$ in $\R ^2$.  
The geometrical role of the Clifford bivectors as rotation operators  will be emphasized below. 

\subsection{The (Pauli) algebra of space} 

From  the usual space $V=\R ^{3}$, with  an ON  basis ($e_{i})_{i=1,2,3}$, we construct $\cliff(\R ^{3}) \equiv  \cliff( 3 )$, the (Pauli) algebra of space. 
Its  elements  are sometimes called the  \blit{Pauli numbers}.

\begin{table}
\begin{tabular}{|c|c|c|c|}
	\hline
	$e_{0}:=1$ & $e_{1}$, ~$e_{2}$, ~$e_{3}$ &$\calI~e_{1}$, $\calI~  
	e_{2} $, $\calI~e_{3}$&$\calI:=\calI~e_{0}$ \\	\hline
	one scalar  & 3 vectors & 3 pseudo-vectors & one pseudo-scalar   \\	\hline
\end{tabular}
	\caption{The basis of the Pauli algebra}
\label{basisPauli}
\end{table} 

{\bf  The orientation operator}

The  antisymmetrical  products of two vectors give three bivectors  (see the table \ref{basisPauli}).
The   trivector  $e_{1}~e_{2}~e_{3} \equiv \calI $ with $ \calI  ^{2}=-1$,   the  \blit{orientation operator},   closes   the  multiplication  law. The   trivectors, also called the     \blit{pseudoscalars}, are all proportional to $\calI$.
They commute  with all elements of $\cliff( 3 )$.  

The \emph{center} of $\cliff( 3 )$, \ie, the set of elements which commute with all elements, is $$\cliff^0( 3 ) \oplus  \cliff^3( 3 ) =   \myspan(1,\calI).$$  The similar algebraic properties of  $\calI$  and  of  the complex pure imaginary $i$   allow  us   to define  an  algebra  isomorphism between $ \myspan (1, \calI ) \subset \cliff(3) $ and $\setC \equiv   \myspan (1,i )$.

 We may write any  bivector $$e_{\mu}~e_{\nu}= e_{\mu}~e_{\nu} e_{\rho}~Êe_{\rho}=\calI~Êe_{\rho},$$ where the index $\rho$ is defined through $ \epsilon_{\mu \nu \rho}=1$. This  allows us to rewrite  the basis of $\cliff(3)$  under the form  $$1,  (e_{i}), (\calI ~e_{i}),\calI$$ (see table \ref{basisPauli}). 
This divides  $\cliff(3)$ into \begin{itemize}
  \item 
a  \guill real "   part $ \cliff ^0(3)\oplus \cliff ^1(3)   \equiv \{paravectors\} \equiv \myspan (1,   e_{i})$: the real multivectors identify to the  paravectors. 
  \item 
 and an \guill  imaginary " part:   $\cliff ^2(3) \oplus \cliff ^3(3)   \equiv \myspan ( \calI~e_{i} ,\calI)$.  
 An imaginary  multivector  is the  sum of a bi-vector  and a tri-vector.
 \end{itemize}
Thus,   any Pauli number may be seen   as a complex paravector. 

{\bf  Pauli algebra and matrices}

With    the identification above (of $\calI$ by $i$), the restriction of the multiplication table (\ref{The multiplication table for Pauli})  to the four paravectors  $(1, e_{i}) $     identifies 
 with that  of    the four Pauli  matrices $(\Id, \sigma _i)_{i=1,2,3}\equiv (\sigma _{\mu})_{\mu=0,1,2,3}$. 
  Thus,     the  real part $\cliff  ^{real}(3)$ is    isomorphic (as an vector space) to $\Herm (2)$, the set of Hermitian complex matrices of order 2. This isomorphism extends to an 
   algebra isomorphism between the complete algebra     $\cliff  (3) $ and    the   algebra of complex matrices of order 2, $M_2(\setC)$,    explicited as 
 \begin{eqnarray}
 1,   ~ e_i &\calI~Êe_i,&\calI   \\
1,    ~\sigma_i &i~Ê\sigma_i,&i   
\end{eqnarray}  
 
The three grade 1 vectors  $e_i$  identify with the 
 three traceless   Hermitian matrices   $ \sigma _i$, which span   $\Herm_0(2)$ (traceless Hermitian matrices).

\begin{table}
 \begin{tabular}{|c|c|c|c|c|c|c|c|}
	\hline
	1 & $e_{1}$& $e_{2}$ &$e_{3}$ &$\calI~ e_{1}$ &$\calI~ e_{2}$ &$\calI~ e_{3}$ &$\calI  $ \\
	\hline
	$e_{1}$ & $1$& $\calI~e_{3}$ &$-\calI~e_{2}$ &$\calI$ & $-e_{3}$ & $e_{2}$ &$\calI~ e_{1}$ \\
	\hline
	$e_{2}$ & $-\calI~e_{3}$&1 &$\calI~e_{1}$ &$e_{3}$ &$\calI $ & $-e_{1}$ &$\calI~e_{2}$ \\
	\hline
	$e_{3}$ & $\calI~e_{2}$& $-\calI~e_{1}$ &1& $-e_{2}$ & $e_{1}$ &$\calI$ &$\calI~ e_{3}  $ \\
	\hline
	$\calI~ e_{1}$  &$\calI$& $-e_{3}$ &$e_{2}$ & $-1 $ & $-\calI~e_{3} $ &$\calI~ e_{2}$&$- e_{1}$  \\
	\hline
	$\calI~ e_{2}$  & $e_{3}$& $\calI$ &$-e_{1}$ &$\calI~ e_{3}$ & $-1$ & $-\calI~e_{1}$&$-e_{2}$  \\
	\hline
	$\calI~ e_{3}$  & $-e_{2}$& $e_{1}$ &$\calI$ &$-\calI~ e_{2}$ &$\calI~ e_{1}$&$-1$&$-e_{3}$\\
	\hline
	$\calI$ & $\calI~e_{1}$& $\calI~e_{2}$ &$\calI~e_{3}$ &$- e_{1}$ & $-e_{2}$ &$-e_{3}$ &$-1$\\
	\hline
\end{tabular}
	\caption{The multiplication table for the Pauli algebra}
\label{The multiplication table for Pauli}
\end{table}

 {\bf  Quaternions in $\cliff(3)$}

It is easy to check, from the multiplication table (\ref{The multiplication table for Pauli}),     the algebra isomorphism 
$$\cliff(3)^{even} \equiv  \setH, $$ the algebra of quaternions.    
Here $\cliff ^{even} (3)$ is the algebra of   even  elements, scalars and bivectors. (However, the odd elements do not form a sub-algebra.) The isomorphism is realized through 
 $$1 \leadsto j_{0}, ~Ê\calI ~e_{1} = e_{2}e_{3} ~Ê\leadsto  j_{1}, ~Ê-\calI ~e_{2} = -e_{3}e_{1} \leadsto j_{2},~  \calI ~e_{3}  = e_{1}e_{2}\leadsto j_{3}.$$  

We may extend   the isomorphism  with $$\calI \leadsto i$$
(the pure imaginary $i$, with the usual rule $i^2=-1$), with  the prescription that $i$  commutes with the four $j_{\mu}$.
This 
 allows us to the $\cliff (3)$ as the set of  \blit{complex quaternions}, 
 $\setH \times \setC$. The complex conjugation $i \mapsto -i$ is distinguished from the quaternionic conjugation $j_i \mapsto - j_i$. It does not reverse the order of the product.

We have also an isomorphism between 
$\cliff ^{even}(3)$ and $\cliff(0,2)$, resulting  from the trivial identification of the latter with the  agebra $\setH$.
  
{\bf Paravectors span \minks}

The paravectors are the Clifford numbers of the form $x=x^0~Ê1+x^i~Êe_i =x^\mu ~Êe_\mu$
 (summation  is assumed  over $i=1,2,3$, and $\mu=0,1,2,3 $).
This allows us to see the  \minks ~as  naturally embedded
in the Clifford algebra of $\R ^3$ \cite{Baylis}, as the vector space of  paravectors, $\clif _0(3)\oplus \clif _1(3)$. 
 
The   [Clifford]Ê conjugation $$x=(x^{\mu})=(x^{0},  x^{i}) \mapsto \bar{x}=(\bar{x}^{\mu}) \equiv (x^{0},  -x^{i}) $$  allows us to  define a quadratic form for the paravectors (which differs, however, from the Clifford scalar product defined above)  
   $$Q(x,y)\equiv \undemi ~Ê( \bar{x} y+ \bar{y} x)=\eta _{\mu\nu}~Êx^\mu~Êy^\nu,$$
where  $\eta$ is the \mink ~norm.

This provides  the  vector space isomorphisms:
\begin{eqnarray}
\clif_0(3)\oplus\clif_1(3)   & \eqsim  M & \eqsim  \Herm (2)\nonumber\\
(\mbox{paravectors})   & \eqsim  (\mbox{\minks})& \eqsim  \Herm (2)\nonumber\\
c=x^0+x^i~e _i                  & \eqsim   (x^\mu)=( x^0,x^i)& \eqsim m=x^\mu~\sigma_\mu \nonumber\\ 
 Q(c,c)  & =  \eta(x,x)& = \det~ m \nonumber\\ 
x^0 & =      x^0  &= \undemi ~\trace ~m. \nonumber 
\end{eqnarray}      
(We have included the isomorphism between \minks ~and Hermitian matrices.) The three grade 1 vectors  $e_i$  identify with  the  three ON  basis vectors   of $\R ^{3}\subset M$.
 Incidently, this suggests that the choice of signature (1,-1,-1,-1) for \minks ~may be more natural that  (-1, 1, 1, 1).
For a development of this approach, see \cite{Baylis}.

\subsection{Spinors}

In the next section, we will introduce the spinors of \spt, and later  we will link them with the \spt ~algebra $\clif(1,3)$.  Here we give some preliminary insights, to show how spinors appear from a purely algebraic point of view (we follow \cite{CAPELAS}). 

First we remark that, in $\cliff (3)$, the two elements (among others) $$e_{\pm} := \frac{1}{2} (1\pm e_3)$$ are idempotent, \ie,  $e_{\pm} ^2=e_{\pm} $. Further, the sets 
$\cliff (3)~e_{\pm} $ and $e_{\pm} ~\cliff (3)$ are left and right ideals of $\cliff (3)$, respectively. They are vector spaces of (complex) dimension 2, and the  identification of  $\calI$ to the complex imaginary $i$  makes each of them  identical to $\setC ^2$. As we will see, a spinor is precisely an element of a two dimensional  representation space for the group SL$ (2,\setC)$, which is 
$\setC ^2$. 

Let us first consider  $\cliff (3)~e_{+} $. If we chose   an arbitrary frame
(for instance   $  \vectord{1}{ 0}=e_{+} ,~ \vectord{0}{ 1} =~e_{1}~Êe_{+}  $ ), we may   decompose an arbitrary element $$ \forall \varphi \in \cliff (3)~e_{+} ,~  \varphi=  \vectord{\varphi ^1}{ \varphi ^2}  \in \setC ^2.$$  We write   $\varphi =(\varphi ^{A})_{A=1,2}$ and $\cliff (3)~e_{+} =\calO ^A$. Such elements constitute  a  representation, called $D^{(1/2,0)} $,  of the special linear group SL$ (2,\setC)$. It corresponds to the so called  \blit{Weyl spinors}, that   we will study in more details below.
 
A  similar procedure applies to   $e_{+}~\cliff (3) $. Choosing a basis (\eg, 
 $ (1, 0) =e_{+}   ,   (0, 1) =~Êe_{+} ~e_{1} ) $, we write its vectors   with covariant (rather than contravariant) and pointed indices:\begin{equation}
\label{ }
 e_{+}~\cliff (3) \equiv \calO_{\dot{A}} = \{\xi \equiv (\xi _{\dot{A}})\equiv (\xi _{\dot{1}},\xi _{\dot{2}})\}.
\end{equation}

We have the very important mapping\begin{eqnarray}
 &  \calO ^A \times   \calO_{\dot{A}}& \mapsto \cliff(3) \\
  & \varphi=(\varphi ^A),\dot{\xi}=(\dot{\xi}_{\dot{A}}) & \mapsto \varphi ~\dot{\xi}=\left( (\varphi ~\dot{\xi})^A _{\dot{A}}=\varphi ^A~ \dot{\xi}_{\dot{A}}\right).\end{eqnarray} The last relation is a matrix product. It provides a complex matrix of order 2, identified to a point of $\cliff (3)$ as indicated above.   

\section{Spinors in \minks}

In the following section, we will study  the \emph{\spt ~algebra}, which is the Clifford algebra of the \minks. This will provide a natural way to consider  the group Spin and the spinors, with the main advantage to allow generalization in any number of dimensions. 
Before turning to the study of the \spt ~algebra,  this section presents an  introduction to the   spinors and the Spin group without 
reference to the   Clifford algebras. The link will be made in the next section. 

We introduce  spinors in \minks ~$\bbbM =\R ^{1,3}$, from group theoretical considerations. 
We recall that the isotropy group of \minks ~is the  \blit{orthogonal group} O$(1,3)$, with four connected components, and which may be seen as a matrix group in its fundamental representation. The restriction to  matrices 
 with determinant 1 leads to the  \blit{special orthogonal group}  SO$(1,3)$, with   2 connected components. Finally, the component of  SO$(1,3)$ connected  to the unity  is the  \blit{proper} Lorentz group $\SOf(1,3)$. None of these groups is singly connected. Their (1-2) \uc s are respectively the groups $\Pin(1,3)$, $\Spin(1,3)$ and  $\Spinf(1,3)$ (see, \eg, \cite{Varlamov}).
 The groups O, SO  and 
$\SOf$ act on \minks ~through the fundamental representation. The construction of spinors is based on the group isomorphisms 
$$\Spinf(1,3)=\SL(2,\setC)=\Sp(2,\setC).$$ (Note also the group isomorphism $\SOf(1,3) = \SO(3,\setC)$).


\subsection{Spinorial coordinates in \minks}
 
 There is a one to one correspondence between the (real)  \minks ~$\bbbM$ and the set $\Herm(2) \subset \Mat_2(\setC)$ of 
{\sl  Hermitian} matrices: to any point $x=(x^{\mu})$ of  $\bbbM$   is associated     the  \blit{Hermitian} matrix  
   \begin{equation}\label{spinrcoord}
  X:=x^{\mu }~\sigma _{\mu}:=\matrixdd{X^{1\dot{1}}}{X^{1\dot{2}}}{X^{2\dot{1}}}{X^{2\dot{2}}}
 := ~ \matrixdd{x^{0}+x^{1}}{x^{2}+i~x^{3}}{x^{2}-i~x^{3}}{x^{0}-x^{1}},\end{equation}
where the $\sigma _{\mu}$ are the Pauli matrices. The matrix coefficients 
 $ X^{A\Adot}$, with  $A=1,2$, $\dot{A}=\dot{1},\dot{2}$, are the  \blit{spinorial coordinates}.
The reason for this appellation will appear below.  We have: \begin{equation}
x \cdot x=\det X,~2~x^{0}=\tr X 
\end{equation} (the dot denotes the scalar product in \minks). In the following, we will distinguish  usual ($x^{\mu}$) and 
spinorial ($x^{A\dot{A}}$) coordinates only  by the indices. Hermiticity reads 
$X^{A\dot{B}}=\overline{X^{B\dot{A}} }$.

An element of the  Lorentz group acts on the \Minkv ~$\bbbM$ as a matrix $L:~x \mapsto L~x$ .
The \emph{same} action is expressed in $\Herm(2)$ through a matrix  $\Lambda$ as 
\begin{equation}
\label{Lorentzaction}
X \mapsto \Lambda ~X~ \Lambda^\dag.
\end{equation}
Here, $\Lambda$ is a matrix of the group
Spin,  the \uc ~of the Lorentz group SO. (When there is no risk of confusion, I will write, \eg, SO for SO(1,3). This will help to recall that most of the derivations below hold in any even dimension, for the Lorentzian case ; odd dimension or other signatures allow similar, although non identical treatments). We have the group homomorphism \begin{eqnarray}
\varphi : & \Spin & \mapsto  \SO \\
  & \Lambda & \mapsto  L.
\end{eqnarray}
To insure the action (\ref{Lorentzaction}), we may chose the matrix $\Lambda$ such that
$$\Lambda ~\sigma_\nu~\Lambda^\dag=L^\mu_{~\nu}~\sigma_\mu,$$which implies 
$$L^\mu_{~\nu}=\undemi~\trace(\Lambda~\sigma _\nu~\Lambda^\dag~\sigma _\mu).$$

Note that $\Lambda$ and $-\Lambda$ correspond to the same element of the Lorentz group, which reflect the fact that $\Spin$ is the 1-2 \uc ~of SO.

***
At the infinitesimal level, $\Lambda\sim \Id +\lambda,~L^\mu_{~\nu}\sim \delta ^\mu_{~\nu}+ \ell^\mu_{~\nu}$, so that 

$$\lambda ~\sigma_\nu~+ ~\sigma_\nu~\lambda^\dag\sim   \ell^\mu_{~\nu}~\sigma_\mu,$$which implies $\lambda =A  \ell^\mu_{~\nu}~\sigma_\mu~\sigma_\nu$.

$$A ~Ê( 2~\ell^0_{~i}~ \sigma_i   
+2\ell^j_{~0}~\sigma_j~ 
 )\sim   \ell^i_{~0}~\sigma_i,$$

which implies $\lambda =A  \ell^\mu_{~\nu}~\sigma_\mu~\sigma_\nu$.
***

\subsubsection{The complex \minks}

The   \blit{complex \minks} ~$\bbbMc$  is defined  by extending the coordinates to 
complex numbers, and extending the \mink ~metric to the corresponding bilinear (not Hermitian)  
form $g(z,z') \equiv \eta _{\mu \nu} ~z^{\mu}~z^{\prime \nu}$.  The same spinorial  correspondance as above leads to identify  $\bbbMc$  with the set $\Mat_2(\setC)$   of all  (not necessarly Hermitian)    \blit{complex  matrices}  $Z=[Z^{A\Adot}]$:  \begin{equation}\label{spinccoord}
\setC ^4 \ni z=(z^{\mu} ) \mapsto  Z \equiv \matrixdd{Z^{1\dot{1}}}{Z^{1\dot{2}}}{Z^{2\dot{1}}}{Z^{2\dot{2}}}
\equiv  \matrixdd{z^{0}+z^{1}}{z^{2}+i~z^{3}}{z^{2}-i~z^{3}}{z^{0}-z^{1}}.
\end{equation}

Any   2 $\times$ 2 matrix with complex coefficients
admits a    unique    decomposition  that we write  $ Z = X+i~Y$,  where 
$X $ and $ Y $  are  both Hermitian:\begin{equation}
{X}=\frac{ Z+ Z^{\dag} }{2},~~~~ Y=i~\frac{ Z- Z ^{\dag} }{2},\end{equation} and $ ^{\dag}$ 
denotes the conjugate transposed.  Hereafter, $iY$ will be called the 
anti-Hermitian part.  
 
 This spinorial notation identifies $\bbbMc$ with $\Mat_2(\setC)$, the set of complex 2$\times$2 matrices, and 
  $\bbbM$ with $\Herm(2)$, the set of complex 2$\times$2 Hermitian  matrices. 
  
  We have defined the   isomorphism $\bbbMc \mapsto \Mat_2(\setC)$  through the Pauli matrices. This is a peculiar choice.
  More generally, it may be  expressed by the   \blit{Infeld-van der Waerden symbols}. We will however consider here only this representation. 
   
  \subsection{The Weyl   Spinor Space}
 
 The Lorentz group plays a fundamental  role in relativistic physics. 
 According to spinorial or twistorial formalisms, even more fundamental is its  \uc, the group $$\Spinf(1,3) \equiv   \SL (2,\setC) =\Sp (2,\setC),$$
  that we will hereafter  simply write  $\Spinf$, when  no confusion is possible. It is at the basis of the spinor formalism. 

In its fundamental representation,  $\SL (2,\setC)$ is  the subgroup  of GL$(2,\setC)$  (the general linear transformations acting on $\setC ^2$) of 
 those matrices whose determinant =1. It has  complex dimension 3 (GL$(2,\setC)$ has complex dimension 4).
 
Thus,   $\Spinf$=SL$(2,\setC)$    acts naturally  on the  vectors of    $\setC ^2$, which are  called  \blit{Weyl  spinors}, or   \blit{chiral spinors}. This is   the so called  $D^{(0,1/2)}$, or \emph{left}, or \emph{negative helicity},  representation.

 As a vector  of the vector space $\setC^2$,  a  Weyl spinor     expands as
 $\xi =  \xi ^A~\spb_A$  (index summation) in   a basis $(\spb_A)=(\spb_1,\spb_2)$.

Thus it appears as a two-component  column vector  $\xi=\vectord{\xi ^1}{\xi ^2} $   and, \bydef, an element of the   group Spin    acts linearly  on it,   as a $2\times 2$  matrix   $\Lambda$  : 
 \begin{eqnarray}\label{act}
\nonumber \Spin: & \setC ^2 & \mapsto \setC ^2\\
\Lambda : &\xi& \mapsto \Lambda ~Ê\xi .
\end{eqnarray}
 
The  set of Weyl spinors,  with this group action,  is  written   $\calO ^A$. A Weyl spinor is written $\xi ^A$. 
 
\subsection{Symplectic form and duality}

Since $\Spinf$=Sp$(2,\setC)$, it may also  be seen as the group of transformation of GL$(2,\setC)$ which preserve a \blit{symplectic form} $\epsilon$ of $\setC ^2$:
  \begin{eqnarray}
\epsilon:&\calO  ^{A}  \times \calO  ^{A} & \mapsto \setC\\
& ~Ê\xi,\zeta &\mapsto \epsilon (\xi,\zeta)  . \end{eqnarray} 

This  gives to    the Weyl-spinor space $\OA$ a \blit{symplectic structure}  $  (\setC ^2,\epsilon)$. Thus, Spin appears as  the symmetry group of the symplectic space $\OA $.  

A frame of   $\calO  ^{A} $ is   {\sl symplectic} iff 
the symplectic form is represented by the matrix     $$  \varepsilon _{AB} =\epsilon (\mu _A,\mu_B) = \matrixdd{0}{1}{-1}{0}.$$ This justifies the notation since, in a symplectic basis, the component $\epsilon   _{AB}= $ identifies with the familiar    \LC ~symbol.
In vector notation, 
  \begin{equation}  \epsilon (\xi,\zeta)   =\xi^T~Ê\epsilon~\zeta    =\varepsilon _{AB}~Ê\xi ^A\zeta ^B=\xi ^1\zeta ^2-\xi ^2\zeta ^1.\end{equation} The antisymmetric  form $\epsilon$    defines an antisymmetric Spin-invariant  scalar product, called the \emph{ symplectic scalar product}. Antisymmetry implies that the \guill symplectic   norm " of any spinor is zero: $ \epsilon (\xi,\xi )=0$. Note that $\epsilon (\xi,\zeta)=0$ implies that $\xi$ is proportional to $\zeta$.

From now, we will only consider   \blit{symplectic frames}, that we also call ON frames, 
 
     The matrix $\epsilon$  is called the \blit{Levi-Civita spinor} (although it is not a Weyl  spinor, but  a spinor in a more general sense that will appear below). Later, we will consider $\epsilon$  as  the spinorial expression of  [the square root of] the \mink ~metric.

\subsubsection{Duality  and the dual representation}

The dual $\calO _A=(\OA)^*$ of  the  vector space $ \OA$  is the space of one-forms on it.  They  are isomorphic.   The symplectic form $\epsilon$  on $\OA$ provides a    \blit{duality isomorphism}  between both spaces: 
\begin{eqnarray}
\epsilon:&   \calO ^A  &\mapsto   \calO _A \equiv (\OA)^*\\
 & \xi  & \mapsto \xi^*=   \epsilon (\xi  , \cdot)  \\
 & \mu _A & \mapsto \mu ^A=   \epsilon (\mu_A, \cdot)  .\end{eqnarray}
 
 To the frame $\spb_A$ is associated the co-frame $\spb ^A$. 
An element of $\calO _A$ expands as  $\eta =\eta _{A}~\spb ^A$, and the symplectic isomorphism is written as  a raising  or lowering of the spinor-indices. Hence the abstract index  notation $   \calO _A \equiv (\OA)^*$.

This is in complete analogy with the metric (musical) isomorphism  defined by a metric in a [pseudo-]Riemannian manifold. Care must be taken however that the calculations differ because of the antisymmetry of $\epsilon$. For instance, we have $u^A ~Êv_A=-u_A~Êv^A$ (sum on indices).

The naturally induced (dual)   action of an element of the  Spin group,
$$\Lambda :~\eta \mapsto \eta ~\Lambda ^{-1}; 
~\eta _A \mapsto \eta _B~  (\Lambda ^{-1})^B_{~A} ,$$ 
  defines  the dual representation, that we note   $\Spin^*$.

 \subsection{Dotted  spinors and the conjugation isomorphism} 
 
    {\bf Complex  conjugation}
    
On the other hand, the (complex)  conjugate    representation  $ \overline{\Spin} $  of the group   Spin  on $\setC ^2$ 
 is defined as \begin{equation}
\label{ }
\Lambda : \eta \mapsto \bar{\Lambda}~\eta ,~\eta \in  \setC ^2 ,
\end{equation}instead of (\ref{act}),     where the bar denotes the complex conjugate.
It preserves also the   symplectic form  $\epsilon$  on $\setC ^2$. We note $ \overline{\OA}  \equiv  \OAdot  $ this representation  vector space. An element is written 
with  \blit{dotted indices}, as  
$ \eta = (\eta ^{\Adot})=\vectord{\eta ^{\undot}}{\eta ^{\deuxdot}}$, where the index $\dotA$ takes the values $\dot{1},\dot{2}$. We call $ \overline{\Spin} $ the group acting in this representation,  the $D^{(1/2,0)}$, or \blit{right representation}.

Complex conjugation   defines the isomorphism  (called \emph{anti-isomorphism})    
\begin{eqnarray}
\OA& \mapsto &\OAdot  \\
\xi = \xi ^A = \vectord{\alpha}{\beta}& \mapsto &  \bar{\xi}= \bar{\xi}^{\Adot}  = \vectord{\bar{\alpha }}{\bar{\beta  }}
.\end{eqnarray} We write $\bar{\xi}$ with dotted indices since it belongs to   $\OAdot$. Since notation may appear confusing,  it is important to remark that $\xi$ and $\bar{\xi}$ are considered as members  of  two different (dotted and undotted) spinor  spaces.

The symplectic structure $  \epsilon$ being preserved by 
the  anti-isomorphism, $\varepsilon _{AB}=\varepsilon _{\dotA\dotB}.$  It also     allows  to raise   or lower the dotted  indices:
  \begin{eqnarray}
\epsilon  :&\calO ^{\dotA } &\mapsto   \calO _{\dotA } \\
 & \zeta & \mapsto   \epsilon  (\zeta, \cdot) \\
 &\zeta ^{\dotA }&\mapsto \zeta _{\dotA }=\zeta ^{\dotB}~\varepsilon _{\dotB \dotA }. \end{eqnarray}  
 
 The symplectic form is also preserved: 
 $$ \Lambda^{-1} ~Ê\epsilon ~Ê\Lambda=\epsilon=\bar{\epsilon}=
\bar{\Lambda}^{-1} ~Ê\epsilon ~Ê\bar{ \Lambda}.$$ \\
 \\
 
  \subsection{ Spinor-tensors and the   \mink ~Êvector space}

The general element  $Z$  of   the tensor product   
$$  \OAAdot  \equiv \OA \otimes \OAdot  $$
 is  called a \blit{mixed  spinor-tensor}  of rank~2. 
In a symplectic  basis, it   expands as $$Z = Z ^{A \Adot}~~\spb _A Ê\otimes   \overline{\spb}_{\Adot} ,$$ and so is represented by the  complex 2$\times$2 matrix $Z  \in   \Mat_2(\setC)$ with components $Z ^{A \Adot} $. Using  the Pauli matrices as  a (complex) basis of $\Mat_2(\setC)$, it expands in turn as $Z=Z^\mu ~Ê\sigma_\mu, Z^\mu \in \setC$. It    identifies  
  with the (complex) vector $ z \in \bbbMc$ with components $z^\mu=Z^\mu=Z^{A\Adot}~(\sigma^\mu)_{A\Adot}$.

The  elements of the form  $Z=\xi~\otimes~ \overline{ \zeta}=\xi ^A~  \overline{ \zeta}^{\Adot}~\spb _A~Ê\otimes   \overline{\spb}_{\Adot} $ are called \blit{decomposable}. 
 In matrix notations, $$Z \equiv  \xi~ \zeta ^T~:~ Z^{A\Adot}  =\xi^A \zeta^{\Adot} ,$$ where the subscript $T$ indicates matrix (or vector) transposition.

This establishes a one-to-one correspondence between \begin{itemize}
  \item vectors $z$ in  \emph{complex}  \Minkvs ~$\Mkc$,  
     $z^\mu=Z^{A\Adot}~(\sigma^\mu)_{A\Adot}$; 
  \item 
      complex 2$\times$2 matrix $Z  \in   \Mat_2(\setC)$ with components $Z ^{A \Adot} $;
        \item 
       \emph{mixed  spinor-tensor}  of rank 2,   
 $Z ^{A \Adot}~~\spb _A Ê\otimes   \overline{\spb}_{\Adot} $.

\end{itemize}    
Spinor-tensors associated to     Hermitian matrices are called \blit{Hermitian} also. They span  the \emph{real} \Minkvs ~$\bbbM$. This is the vector, or $(\undemi,\undemi)$, representation.

For decomposable spinor-tensors, the \blit{scalar product} is defined through the symplectic form, as
$$\eta(\xi~\otimes~ \overline{ \zeta}, \xi'~\otimes~ \overline{ \zeta}')=\epsilon(\xi,\xi')~
\epsilon(\zeta,\zeta'),$$ and extended by linearity. It is easy to check that it coincides with the \Mink ~norm for the Hermitian spinors. 

A \blit{decomposable}    spinor-tensor $Z^{A\Adot}=\xi  ^A~ \bar{ \zeta}^{\dotA  } $  corresponds to a   \blit{null} vector (of zero norm)    in $\bbbMc$, not necessarily real.
Those of the form     $Z^{A\Adot}=\xi ^A ~\bar{\xi}^{\dotA }=$ are Hermitian ($Z^{A\Adot}= \overline{ Z}^{ \Adot A}$) and, thus,   correspond to    null vectors in the \emph{real} \Minks ~$\bbbM$ : they belong to its null cone.

To any Weyl spinor $\xi$ is associated  the  null vector $\xi \otimes \bar{\xi}$  in real \Minks, called its   \blit{flagpole}.     Changing  the spinor phase (multiplying it   by a   complex unit number)  does not change the null vector. Multiplying the spinor  by a real number
multiplies the null vector by the same number squared.
 Note that a null  vector  of \minks ~Êmay be seen as the momentum of a zero mass particle.

The   table (\ref{tablespinor}) summarizes the properties of the  spinor representations.

\begin{table}	
\begin{tabular}{|c|c|c|c|c|c|}
	\hline
Space   & ¥ & indexed  spinor   & Representation&form \\
   &   &     &   &  \\
	\hline
	\hline
Weyl spinors  & $ \calO ^{A}$ & $ f^{A} $ & $	f^{A} \mapsto \Lambda ^A_{~B} ~Êf^{B}$ & $\epsilon _{AB} $\\
($d_\setC=2$) & ¥ & ¥ & ¥ &  ¥  \\
	\hline
\emph{dual}	¥ & ¥ & ¥ & ¥ & ¥  \\
Weyl spinors & $(\OA)^* =\calO _{A}$ & $f_{ A} $ &  $	f_{A} \mapsto (\Lambda ^{-1})_{~A}^B ~Êf_{B}$ & $\epsilon  ^{\Adot \Bdot}$\\
	\hline
\emph{dotted } & ¥ & ¥ & ¥ & ¥  \\
Weyl  spinors & $\overline{ \OA}=\calO ^{\Adot}$ & $  f^{\Adot} $ & $	f^{\dotA} \mapsto \bar{\Lambda} ^{\dotA} _{~\dotB} ~Êf^{\dotB}$ &$ \epsilon _{\Adot \Bdot}$\\
	\hline
	dual dotted & ¥ & ¥ & ¥ & ¥  \\
Weyl spinors   & $ \overline{ \OA}^* =\calO _{\Adot}$ & $  f_{\Adot} $ & $	f_{\dotA} \mapsto (\bar{\Lambda} ^{-1}) _{~\dotA} ^{\dotB} ~Êf_{\dotB}$ & $\epsilon ^{AB}$\\
 ($d_\setC=2$)& ¥ & ¥ & ¥ & ¥¥  \\
	\hline 
	\hline 
Complex	  & $\calO ^{A \Adot} =\calO ^{A} \otimes \calO ^{\Adot}$&$  f^{A\Adot}   $ & $f^{A\Adot}  \mapsto \Lambda ^A_{~B}~\bar{\Lambda} ^{\dotA} _{~\dotB}~f^{B\Bdot}  $ &$ \epsilon _{AB} ~\epsilon _{\Adot \Bdot} $\\
\mink   &   $\simeq \Mkc $& $ \simeq f^a $  & $f^a  \mapsto  L^a_{~b}~f^b$ & $\simeq \eta _{ab}$  \\
\spt       &    & ¥ & ¥  & ¥  \\
 ($d_\setC=4$)   &    & ¥ & ¥  & ¥  \\
\hline 
 	dual~  &$ \calO _{A \Adot} =\calO _{A} \times \calO _{\Adot} $ &$ f_{A \Adot}  $ & $ f_{A \Adot} \mapsto   $ &  $\epsilon ^{AB} ~\epsilon ^{\Adot \Bdot} $\\
             &  &  & $  
	  (\Lambda^{-1}) _A^{~B}~ (\bar{\Lambda}^{-1}) _{\dotA} ^{~\dotB}~f_{B\Bdot}  $ &  \\
       & $\simeq \bbbMc^{*}   $ & $ \simeq f_{a}    $  & $f_{a}  \mapsto L_a^{~b}~f^b$  & $\simeq \eta^{ab}$ \\
\hline 
	\hline 
\emph{real }	\mink   & $ (\calO ^{A} \otimes \calO ^{\Adot})_{Herm.}$&$f^{A \Adot}   $ {Herm.}  &  & ¥  \\
 	\spt   &$\simeq \Mk$  & $ \simeq f^a $ ~real &$f^a  \mapsto  L^a_{~b}~f^b$  & $\eta_{ab}$\\
 ($d_\R=4$)   &    & ¥ & ¥  & ¥  \\
	\hline 
 	dual   ~($d_\R=4$)& $ (\calO _{A} \times \calO _{\Adot})_{Herm.}$	&$ f_{A \Adot} $ {Herm.}   &  & ¥  \\
 	    & $\simeq \bbbM^{*} $	&$\simeq  f_{a}  $ real & $f_{a}  \mapsto L_a^{~b}~f^b$  & $\eta^{ab}$ \\
	\hline 
	\hline \end{tabular}
\caption{Spinor vector spaces with  their tensor products, and their links with \minks.  }
\label{tablespinor}
	\end{table}	

{\bf Two-component spinor calculus in \minks}
 
The use of  spinorial indices in \minks ~Êmay be seen as a simple change of notation: each tensorial index is replaced by a pair $A\Adot $ of spinorial indices and all usual formulae of tensorial calculus hold. For instance,   the gradient    $\nabla _\mu$ becomes    $\nabla _{A\Adot}:=\pder{ }{X^{A\Adot} }$ and, for 
any function $f$, $\d f= \nabla _{A\Adot}f  ~\d X^{A\Adot} $.

Similarly, tensors in \minks ~appear with  spinorial 
indices, like $S^{AB\ldots \Adot \Bdot  \ldots}$. When the dotted and undotted  indices appear in pairs, the tensor-spinor may be seen also as   a tensor over  \minks, written in spinorial notations.  Quite often, one needs the  symmetrized or antisymmetrized combinations of indices, of 
the types $S^{(AB\ldots)\Adot \Bdot \ldots}$ or $S^{[AB\ldots ] \Adot \Bdot \ldots}$, 
etc.
Any form or  tensor in \minks ~can  be written in spinorial  components. In particular,  the \mink ~metric,
\begin{equation}
\eta _{\mu \nu}~\sigma _{A \Adot}^{\mu}~\sigma _{B\Bdot}^{\nu}=\varepsilon 
_{AB}~\varepsilon _{ \Adot\Bdot}. \end{equation}
One  simply writes usually
$$\eta _{\mu \nu} \simeq  \varepsilon _{AB}~   \varepsilon _{ \Adot\Bdot} $$
 
{\bf Spin group and Lorentz group}

The      action of  the group Spin on $\OA$ and $  \overline{\OA}$  induces the following  action on the  tensorial product $\OAAdot $, 
\begin{eqnarray}
\mbox{Spin}: & \OAAdot  = \Mat_2(\setC) \mapsto &  \OAAdot =  \Mat_2(\setC) \\
\Lambda:& Z=\xi ~ \zeta  ^{ T} \mapsto & (\Lambda \xi)~(\bar{\Lambda} \zeta)^{T}= \Lambda ~Z~\Lambda ^{\dag}\nonumber.\end{eqnarray}   
 We recognize      the action (\ref{Lorentzaction}) of the Lorentz    group   on $\Mat_2(\setC) = \bbbMc$.
Thus the mixed spinor-tensors of $\OAAdot $  are really vectors of  the (complex) \minks: $$(\R ^{1,3},\eta)\equiv \bbbMc \equiv (\OA,\epsilon) \otimes (  \overline{\OA},\bar{\epsilon}) ,$$
considered as a representation space for the Lorentz group.

Note that the correspondence between $\Mat_2(\setC)$ and $   \bbbMc $ is defined through  the Pauli  matrices.
The reduction to the set  $\Herm_2(\setC) \subset \Mat_2(\setC)$
is the \emph{decomplexification}  of $  \bbbMc$ to the usual \Minkv ~$  \bbbM$. The usual action of the Lorentz group  results.

The correspondence between spinorial and tensorial indice may be seen very simply as replacing any index $\mu$ by a pair $A \Adot$, and conversely. More rigorously, it is expressed by the Infeld-van der Waerden symbols.

 \subsection{Dirac spinors and   Dirac matrices}

A \blit{Dirac spinor}  is constructed as the direct sum of a left Weyl spinor and a right   Weyl spinor. It is written as $$\Psi  =\vectord{\pi }{\eta }=\vectord{\pi ^A}{\eta ^{\Adot}}$$ (here written in the Weyl representation).  We write $S_{Dirac}$ the vector space of the Dirac spinors.
Thus the action of a Lorentz  matrix  is 
defined as : $$\Lambda: ~\vectord{\pi }{\eta }\mapsto \matrixdd{\Lambda}{0}{0}{ \overline{ \Lambda }}~\vectord{\pi }{\eta }.$$

 It is a calculatory exercise to show that the symplectic forms of the Weyl spinor spaces induce an Hermitian metric $\varphi$ of signature 2,2  for the Dirac spinor space. This makes the space of  Dirac spinors appear  as the   fundamental representation of the group  $\SUdd= \Spin(2,4)$. Note that this group is  embedded in $\cliff(\bbbM)$ and that we will consider below the Dirac vector space as a representation space for $\cliff(\bbbM)$. 

{\bf  Dirac matrices}

The \blit{Dirac matrices}   are   four matrices $\gamma _\mu$   acting on $S_{Dirac}$,   obeying the anticommutation relations (see below for the  link with Clifford algebras):\begin{equation}
\label{ }
[\gamma _{\mu}, \gamma _{\nu}]_+= 2 ~\eta _{\mu \nu}.
\end{equation}  
Their indices are lowered or raised with the \mink ~Êmetric. 
Thus,  $$\gamma ^{0}=\gamma _{0},~\gamma ^{i}=-\gamma _{i}.$$

{\bf  Chirality}

One defines the \blit{orientation operator} $\gamma _0~\gamma_1~\gamma_2~\gamma_3$, of square $-\Id$, and the   \emph{chirality operator}
 $$\chi=\frac{i}{4}~\epsilon_{\mu\nu\rho\sigma}~\gamma ^\mu~\gamma^\nu~\gamma^\rho~\gamma^\sigma=i~\gamma ^0~\gamma^1~\gamma^2~\gamma^3~.$$
($-\chi$ is also written $\gamma_5$.)

The Weyl spinors may be seen as the  eigenstates of $\chi$, with eigenvalue $\pm 1$, in $S_{Dirac}$.
The projection operators $\undemi (\I \pm \chi)$ project a Dirac spinor into a left or right spinor.
So that a general Dirac spinor may be written $$\Psi =\vectord{\psi _R}{\psi _L},$$ involving    the left and right  Weyl representations. 

 The parity transformation is defined as 
  $$\Psi \mapsto \gamma ^0 ~Ê\Psi.$$

The \blit{rotation generators} are
$$\Sigma ^{mn}=-\unquart [\gamma ^{m},\gamma^{ n}].$$

$$Ê\psi^\mu \mapsto  - \undemi (\Sigma ^{mn})^\mu_\nu~Ê\psi^\nu.$$

{\bf Charge conjugation and Majorana spinors}

The  \blit{charge conjugation} is defined as the operation 
$$ \Psi \mapsto \Psi ^c = - \gamma ^2 ~Ê\overline{\Psi} ,$$
where the bar means complex conjugation.
From the physical point of view, 
the charge conjugation transforms a particle into an antiparticle.

The \blit{Majorana spinors} are defined as those Dirac   spinors
which are  self-conjugate under charge conjugation : 
$$\Psi =\Psi ^c.$$
The space of 
Majorana spinors has complex dimension 2.

{\bf Representations}

There are different representations, depending on the basis in which they are written. In the \blit{Weyl representation} (or  \blit{chiral representation}), we have
\begin{equation}\label{ }
\gamma ^{0}=\matrixdd{0}{\I _2}{\I _2}{0},~
\gamma ^{i}=\matrixdd{0}{-\sigma ^{i}}{ \sigma ^{i}}{0},~
Ê\gamma ^{5}=\matrixdd{\I _2}{0}{0}{-\I _2}
\end{equation}

\section{Spin  and spinors in Clifford algebras}

\subsection{Rotations in a vector space}

We first recall some properties of the orthogonal groups.

In the real vector space $V=\R^{p,q}$ with a scalar product of signature $(p,q)$, the group of isometries preserving the scalar product is the orthogonal group O$(p,q)$. For an Euclidean (resp. Lorentzian) signature, it has two (resp. 4)  connected components. Given a basis for $V$, O$(p,q)$ may be seen as a group of matrices. The subgroup   SO$(p,q)$ of matrices with unit determinant has one (resp. 2)  connected components. In the Lorentzian case, one also defines $\SOf $ (also written 
$\SO _0(p,q)$), the component of  SO$(p,q)$ connected to the identity. This is the group of \emph{proper orthogonal transformations}.
For \spt, SO$(1,3)$ and 
$\SOf(1,3)$ are the Lorentz group and the special Lorentz group, respectively. We will study them in more details below.

As we will see, the (2 to 1) \uc ~of  O is the group Pin.
The (2 to 1) \uc ~of SO is the group Spin.
The (2 to 1) \uc ~of $\SOf$ is the group $\Spinf$:
$${\rm O}=\Pin/\setZ_2,$$
$$\SO=\Spin/\setZ_2,$$
$$\SOf=\Spinf/\setZ_2.$$

\subsection{The Clifford group}

A  Clifford algebra is not, in general, a multiplicative group, since  some  elements are not invertible. 
This is for instance the case 
of  the null vectors of the \mink ~vector space   $\Cliff^1({1,3})=\R^{1,3}$. However, we will extract some multiplicative groups from a Clifford algebra, after selecting the invertible  elements.

Given an invertible element $x$ of a Clifford algebra $C=\clif(V)$, we define
its action on $C$ as
\begin{eqnarray}
\nonumber T _x:&C &\mapsto C:\\
&c& \mapsto  -x~c~x^{-1}=\forall c \in C.\end{eqnarray}
It is convenient to demand that this preserves  the vector space $V=C^1$, \ie, 
$$Êv \in V   \implies  x~ v ~x ^{-1} \in V .$$

This  defines the  \blit{Clifford group}, also called \emph{Lipschitz group}, of 
$\clif(V)$ as its subset
\begin{equation} \label{ }  \Gamma  (V) \equiv ~\{x\in  
\clif(V);~x \mbox{~Êinversible~Ê};  ~Êv \in V   \implies x ~ v ~x ^{-1} \in   \clif^1 (V)=V\}.\end{equation}  
The composition law of the Clifford group is  the Clifford multiplication.
 
The  subset    $\Gamma ^{even}$ (also written   $\Gamma ^{+}$) of all \emph{even}  elements of $\Gamma$ is called the \emph{special Lipschitz group};  the  subset of all odd elements  (not a group) is written  $\Gamma ^{odd}$ (or $\Gamma ^{-}$). One also distinguishes 
$ _{\pm}Ê\Gamma $ the set of elements of norm $\pm1$. 

\subsection{Reflections, rotations  and Clifford algebras}

Let us  examine this action when both $c$ and $x$ belong  to  $V \equiv C^1   \subset C   $:
 $$ v  \mapsto  -x~v~x^{-1}=v - 2~ \frac{v\cdot x}{x\cdot x}~x  ,$$with    Clifford products  in the LHS and scalar products in the RHS.  Geometrically, we recognise a  \blit{reflection} in $V$ with respect to the hyperplane orthogonal to $x$.
But we know that any rotation in  $V$  can be  written  as the  product of two  reflections. Thus, the  action of a \blit{rotation}  takes the form :\begin{equation}
T_R \equiv T_v \circ T_w :~Êx \mapsto vw~x~w^{-1}v^{-1}:=R~x~ R^{-1}; ~Êx \in C^1.
\end{equation}  Being  a product of two 1-vectors,     $R$ is   the sum of a scalar plus  a  bivector.  The formula implies that the composition of rotations is represented by the Clifford product  
   (a Clifford  product of even multivectors is an even multivector.)
It is clear, however, that $R$ and $a~R$, with $a$ a scalar, represent the same rotation. Thus, it is natural to introduce a normalisation. Finally, it can be shown  that the  general  rotation in $V$ is  represented, in this way,  by an   even grade multivector   $R$,  such that 
$$R  ~v~Ê R^{-1}\in C_1,~Ê\forall v \in C_1 \andy \bar{R} R =\pm 1.$$ 
This action works on  all  objects in $C$. For instance, their action on the bivectors  is given by  \begin{equation}
T_R:~xy \mapsto R~x~ {R}^{-1} R~y~ {R}^{-1}=R~xy~ {R}^{-1}.
\end{equation}

If we impose $\bar{R} R =+1 $, such an element is called a     \blit{rotor}. The rotors form the rotor  group   $\Spinf$.

\subsubsection{The Pin and Spin groups}
 
 In   the Clifford algebra $\clif(V)$, we define the following subgroups of the Clifford group:

The set of  all (Clifford) products of non null normed  vectors of $V$ ($\bar{s}s=\pm 1$)   form   a group  for the [Clifford] multiplication: this is the \blit{Pin   group}  associated to $V$:\begin{equation}
\label{ }
\Pin (V) \equiv \{ s \in \clif(V);~ s=s_1 s_2 ... s_k, ~Ês_i \in V\equiv \clif_1 (V), ~Ê\bar{s}s=\pm 1  \}.
\end{equation} 
Note that an invertible normalized element of $V$ belongs to $\Pin$.

The  restriction to   the product of even numbers of vectors  gives the    \blit{Spin group} \begin{equation}
\label{ }
\Spin (V) \equiv \{ s \in \clif(V);~ s=s_1 s_2 ... s_{2k}, ~Ês_i \in V \equiv \clif_1 (V), ~Ê\bar{s}s=\pm 1  \}.
\end{equation} 
Thus, $\Spin (V)$ is a subgroup of $\Pin (V)$, and also of  $\clif^{even}(V)$. It  may also be defined as the group of all elements $s$ of C such that $$svs^{-1} \in C_1,~\forall v \in C_1 \andy \bar{s}s =\pm 1.$$
Thus,  $\Spin (V)= \Pin (V)~\bigcap ~ \Cliff^{even}(V)$.

We give in the table \ref{ClgroupT} the relations between the different groups introduced.\\
\\
\begin{table}
\begin{tabular}{|c|c|c| }\hline
Lipschitz group = Clifford group      &$\Gamma$    &\\ \hline
special Lipschitz group      &   $\Gamma ^{even}  $&\\ \hline
Pin group      &   $ _\pm \Gamma  $&u.c. of ~ O(3,1) \\ \hline
Spin  group      &   $_\pm \Gamma ^{even}=$Pin $\cap  ~\Gamma^+$&u.c. of ~SO(3,1) \\ \hline
Rotors   group = $\Spinf $     &   $_+ \Gamma ^{even}=\SL(2,\setC)$&u.c. of~ $\SOf$(3,1).\\
\hline
\end{tabular}\\
\caption{The groups included in a Clifford algebra, as   \uc s for orthogonal  groups}
\label{ClgroupT}
\end{table}

Pin and Spin are double cover representations (\uc s)  of O and SO:   
$${\rm O}=\Pin/\setZ_2,$$
$$\SO=\Spin/\setZ_2,$$
$$\SOf =\Spinf/\setZ_2,$$
  
  $\Spin (p,q)$ and   $\Spin (q,p )$ are isomorphic but this relation does not  hold for the  Pin  groups.
Both are Lie groups. One denotes $\Spin(p)=\Spin(0,p)=\Spin(p,0)$.

\subsubsection{The Clifford - Lie algebra}

Let us define the  \blit{Clifford bracket} as the commutator
$$[x,y]_{Clifford}\equiv \undemi~Ê(xy-yx),$$ where the products on the RHS are Clifford products. 
This provides to $C$ a Lie algebra structure (note that its dimension is one unit less that the Clifford algebra). 

Although the Clifford product of a bivector by a bivector is not, in general a pure bivector, it turns out that the commutator preserves the set  of bivectors. Thus, the set of bivectors, with the   Clifford bracket is the  sub - Lie algebra   $$\myspan  (e_\mu~ e_\nu)_{1\le \mu<\nu \le n},~e_\mu \in \Cliff^1.$$    This is the   Lie algebra of  the rotor group $\Spinf  $,  also of  the groups $\Spin  $  and  SO.

Rotors act  on bivectors through  the adjoint representation of the rotor group,  $T_R : ~ÊB \mapsto R B ~{R}^T = Ad_R (B)$.  
 
 Now we consider all these notions in more details, in the case of the \spt ~Êalgebra. 
 For their extension to an arbitrary number of dimensions, see \cite{Rausch}, \cite{Coquereaux}.

{\bf Examples}

\begin{itemize}
  \item
For the \emph{plane}, the  general rotor is \begin{equation} \cos  \frac{\theta }{2}+\calI_2~\sin \frac{\theta }{2}:=\exp[\calI_2~\frac{\theta }{2}]. \end{equation}
The exponential notation results immediately from its series development, and from the  anticommutation properties of the algebra. The rotation angle $\theta$  parametrizes the rotor group SU(1) = Spin(1), and 
  $u$ is  a spatial   unit vector, 
the axis of the rotation. 

  \item 
For the \emph{space} $\R ^{3}$, the general rotor is of the form  \begin{equation} 
\cos \frac{\theta }{2}+\calI_3~u~\sin \frac{\theta }{2}:=\exp[\calI_3~u~\frac{\theta }{2}]; ~Êu\in \cliff(3)^1,~u~u=1.
\end{equation} The unit spatial vector $u$ is   the  rotation  axis  and $\theta$ the rotation angle. The orientation  of $u$, and the angle $\theta$, parametrize the group SU(2).

  \item 
In $\cliff(1,3)$,  the Clifford algebra of  \Minkv,  we have $ (\calI_4)^T=\calI_4$, and  $ {B}^T= -B$ for an arbitrary bivector $B$. The  general rotor is of the form $\alpha +\beta ~B+\gamma ~\calI$, with $B$ an 
arbitrary bivector which verifies the condition $\alpha ^{2}+\beta ^{2}~B~{B}^T-\gamma 
^{2}+2\alpha ~\gamma ~\calI=1$.

A space+time splitting  allows us to write  the  bivector basis as $(\Sigma _{i}, \calI~\Sigma _{i})_{i=1..3}$, see below. Then, the  general rotor is  written under the form  
\begin{equation}
(\cosh \frac{\phi }{2}+\sinh \frac{\phi }{2}	~u^{i}~\Sigma _{i})~
(\cos \frac{\psi }{2}+\sin  \frac{\psi }{2}~n^{j}~I~\Sigma _{j}) =e^{\frac{\phi }{2}~u^{i}~\Sigma 
_{i}}~e^{\frac{\psi }{2}	~n^{j}~I~\Sigma _{j}} ,
\end{equation}
or
\begin{equation}
(\cos \frac{\psi ' }{2} +\sin\frac{\psi ' }{2}	~n^{\prime j}~\calI~\Sigma _{j}) 
(\cosh \frac{\phi ' }{2}+\sinh \frac{\phi ' }{2}	~u^{\prime i}~\Sigma _{i})~ =e^{\frac{\psi ' }{2}	~n^{\prime j}~\calI~\Sigma _{j}} ~e^{\frac{\phi '}{2}~u^{\prime i}~\Sigma _{i}}.\end{equation}
Care must be taken of the non commutativity on calculations.
A rotor $e^{\frac{\phi   }{2}~u^{i}~\Sigma 
_{i}}$ corresponds to a spatial rotation.
A rotor $e^{\frac{\psi  }{2}~n^{j}~\calI~\Sigma _{j}}$ corresponds to a boost.
\end{itemize}

 \subsection{The \spt ~algebra}\label{sptalgebra}

The  \blit{ \spt ~algebra}  $\Cliff(\bbbM)=\Cliff(1,3)$ 
is the real Clifford algebra 
of \Minkv ~$M\equiv \R^{1,3}$. It has a (real) dimension 16.
Although   it is sometimes called  the  Dirac algebra, we  reserve here  the appellation   to the complex Clifford algebra $\ccliff(4)$ (see below). The latter is the  common complex extension of 
$\Cliff(1,3) \approx \Mat (2,\setH)$ and  $\Cliff(3,1) \approx \Mat (4,\R)$.

The \spt ~algebra $\Cliff(1,3)$  is   generated by  four vectors  $e _{\mu} $ which form an  ON  basis of $\bbbM$:   $e _{\mu} \cdot  e _{\nu}=\eta _{\mu \nu}$.
These four vector, with  their [Clifford]Ê products, induce an ON  basis  of 16 elements for 
$C\ell(\bbbM)$, given in the table (\ref{Diracalgebrabasis3}):  unity 
(scalar of grade 0), 4 vectors (grade 1), 6 bivectors (grade 2), 4 
trivectors (grade 3) and the pseudo scalar $\calI \equiv  e_5 \equiv e _{0}~ e _{1}~ e _{2}~ e _{3}$ (grade 4).  

{\bf Duality}

Calculations show  that $\calI^{2}=-1$, and that $\calI$ anticommutes with the $e_\mu$. The multiplication by $\calI$     exchanges the  grades $r$ and $4-r$. This allows to chose a basis for  the trivectors, under the form of  the four  $\calI~ e _{\mu}$, that we call  \blit{ pseudovectors}. This also provides   a convenient  (altough non covariant) basis for  the bivectors:  after having selected a timelike  (arbitrary) direction $e_{0}$: we define the three time-like bivectors $\Sigma 
_{i} \eqbydef e_{i}~e_{0}$.
Then, the basis is completed by the three $\calI~\Sigma _{i} =e_{j}~e_{k}$.  Note that $\calI=e_{0}~e_{1}~e_{2}~e_{3}=~\Sigma _{1}~\Sigma _{2}~\Sigma 
_{3}$.  The basis is given in table (\ref{Diracalgebrabasis3}).

{\bf Even part}

The even part $\Cliff^{even}(1,3)$   is the algebra   generated by 1,  $\calI$ and the 6 
bivectors $(\Sigma _{i} , ~ \calI~\Sigma _{i})$. It  is isomorphic to  the (Pauli)   algebra  $\Cliff(3)$, generated by the $\Sigma 
_{i}$. The latter generate    the 3-dimensional  space orthogonal to the time 
direction $e_{0}$  in \minks. 
Note that this space-time splitting, which requires the choice of an arbitrary time direction,  is  non covariant.
 
 {\bf Real matrices}
 
Note that $\cliff(\bbbM)$ is isomorphic to $M_4(\R)$, the set of real-valued matrices of order 4.
An  isomorphism may be constructed from an ON  basis $e$ of \minks: first one defines $e^{02} \equiv e^0 e^2$ and the elements \\
$P_{1} \equiv \frac{1}{4} ~Ê(1+e_1)~Ê(1+e_0~Êe_2)$,\\
$P_{2} \equiv \frac{1}{4} ~Ê(1+e_1)~Ê(1-e_0~Êe_2)$,\\
$P_{3} \equiv \frac{1}{4} ~Ê(1-e_1)~Ê(1+e_0~Êe_2)$,\\
$P_{4} \equiv \frac{1}{4} ~Ê(1-e_1)~Ê(1-e_0~Êe_2)$.\\
They verify $P_i~P_j=\delta _{ij}$.

\begin{table}
	\caption{The basis of the \spt ~algebra, $i=1,2,3$, $\mu=0,1,2,3$}
\label{Diracalgebrabasis3}
 \begin{tabular}{|c|c|c|c|c|}
 \hline
	1 & $(e_{\mu})$ &$(e_{\mu}~Ê e_{\nu})$&$ (e_{\mu}~e_{\nu}~Êe_{\rho})$=&$ (e_{\mu}~e_{\nu}~Êe_{\rho}~Êe_{\sigma})$\\
	1 & $(e_{\mu})$ &$(\Sigma _{i}) $, $(\calI~\Sigma _{i})$&$\calI~e_{\mu}$&$\calI = e_5$\\	\hline
	one scalar  & 4 vectors & 6 bivectors & 4 trivectors & one quadrivector   \\	\hline
   &   &   & = pseudovectors & = pseudoscalar   \\	\hline
\end{tabular}
\end{table} 

\subsection{Rotations in \minks}
 
Given an arbitrary invertible    multivector $\lambda \in C\ell(\bbbM)$, we define its  action $T_\lambda$    on $C\ell(\bbbM)$  as  \begin{eqnarray}
T_\lambda: & C\ell(\bbbM) \mapsto & C\ell(\bbbM) \nonumber \\
 &  v~Ê \mapsto & \lambda ~v~ \lambda ^{-1}.
 \end{eqnarray}
Those elements which preserve the \minks ~$\bbbM \equiv C\ell ^1(\bbbM)$
 form  the \blit{Clifford group} \begin{equation}
\label{ }
G_{1,3}=\{\lambda ; ~v \in \bbbM \imply  T_\lambda v \equiv \lambda v \lambda ^{-1} \in \bbbM \}.
\end{equation} 

To  each $\lambda$ corresponds  an element of the group O(1,3) such that  $\lambda v \lambda ^{-1}$ is the transformed of $v$ by its element.
For instance,  when   $\lambda =\pm e_{0}$, $T_\lambda$   represents a space reflection. When
  $\lambda = \pm e_{1}~e_{2}~e_{3}$, $T_\lambda$   represents a time reflection. 
Conversely, any Lorentz  rotation can be written as  $T_\lambda$ for some  (Clifford) product   $\lambda$ of non isotropic vectors. This expresses the fact that   any Lorentz rotation   can be obtained as a product of reflections. 
  
  {\bf   Pin and Spin} 
  
This group action on $\bbbM$  is however not \emph{effective}: the Clifford group is \guill too big ".  One considers the subgroup $\Pin(1,3)$ of $G_{1,3}$ as those elements  which are products  of  unit elements of $\bbbM$  only, \ie, such that  $v \cdot v=\pm 1$. The even part of  $\Pin(1,3)$ is the spin group $\Spin(1,3)$. 
It is multiconnected; its component of the unity is the (2-fold) \uc ~Êof the \emph{proper} Lorentz group,   $\Spinf(1,3)=\Spin(1,3) /  Z_2$.

Finally,   any Lorentz rotation is represented  as   a bivector written   $R=A+\calI~B$ in the basis $(\Sigma  _{i},\calI~\Sigma  _{i})$ above.

We have the Lie group  homomorphism  \begin{eqnarray}
{\cal H}:& \Spin(4)& \mapsto \bbbL =\SO \\
 & \lambda &
\mapsto [\tensor{a}{\alpha}{~\beta}], \nonumber \end{eqnarray}
such that $\lambda ~e_{\alpha}~ 
\lambda ^{-1} = \tensor{a}{\beta}{~\alpha}~e_{\beta} $. Since   
$\lambda$ and $-\lambda$ correspond to the same rotation, ${\cal H}$ is a   2-1~homomorphism.   

More generally, in the Lorentzian case, a special orhogonal group is  not simply connected 
and its  \uc  ~group is   precisely  the spin group. The kernel of the homomorphism $\Spin (1,d-1) \mapsto  $SO$(1,d-1)$ is isomorphic to $\bbbZ _{2}$, so that $\Spin(1,d-1) / \bbbZ _{2} 
\approx  $SO$(1,d-1)$.  $\Spin(1,d-1)$ is the \uc ~of SO$(1,d-1)$.
  
  \subsection{The Dirac algebra and its matrix representations }

  \subsubsection{Dirac spinors and Dirac matrices}

We follow    \cite{Rausch} and  \cite{Coquereaux}.
 The \blit{Dirac algebra} is defined as  the complexification $\cclif(4) = \clif(M) \otimes \setC$ of the \spt ~algebra. This is the Clifford algebra of the [complex]Êvector space $\setC ^4$ with the quadratic form $g(v,w)=\delta_{\mu\nu}~Êv^\mu~Êw^\nu$.
 
 The discussion here is presented for the case of dimension 4. It generalizes to any \blit{even dimension}. 
 In the case of odd dimensions, the things are slightly different, see \cite{Avramidi}.
  
There exists  \cite{Avramidi}    a    (complex) faithful irreductible   representation of the algebra $\cclif (4)$ as $\End (S_{Dirac})$, the group of \blit{linear endomorphisms} of a  vector   space    $ S=S_{Dirac}$, of complex  dimension 4,   the vector space of \blit{Dirac spinors}.
(For a   dimension $n$,  the vector space of  Dirac spinors is of dimension $2^{[\frac{n}{2}]}$. The representation is faithful when $n$ is even.)
  This means    an \blit{algebra    isomorphism}
  $$ \cclif (4) \to \End(S_{Dirac})= \Mat_4(\setC)$$    
  $$ e_\mu \to   \Sigma _\mu.$$    
Since $S_{Dirac}$ is a complex vector space of complex dimension 4, 
$\End(S_{Dirac})$  is  isomorphic to $\Mat_4(\setC)$.

The unity is represented by identity; the  four   1-vectors  $e_\mu$ of $\ccliff ^1  $ are   represented  by four complex $4 \times 4$    matrices  $(\Sigma _\mu)_{\mu=0,1,2,3}$, which verify the Clifford  anticommutation relations \begin{equation} \label{ }
[\Sigma _{\mu}, \Sigma _{\nu}]_+=2 ~\delta_{\mu \nu}.\end{equation}


The \blit{ \spt ~algebra}  $\cliff(1,3)$ is the [real]Êsub--algebra of  $\cclif (4)$ generated by the four elements $e_0, ie_i$. 
Thus, the previous representation of $\cclif (4)$ 
 provides  a matrix representation of $\cliff(1,3)$, obtained   by defining the four \blit{Dirac matrices} $(\gamma _\mu)_{\mu=0,1,2,3}$:
$$\gamma_0=\Sigma _0,~\gamma_i=i~\Sigma _i.$$
They represent the four   1-vectors  $e_\mu$ of \Minkvs.
The other representatives are found by explicitation of the products.
 They    verify the Clifford  anticommutation relations \begin{equation} \label{ }
[\gamma _{\mu}, \gamma _{\nu}]_+=2 ~\eta_{\mu \nu}.\end{equation} These matrices (and  their  products) act as operators  on the vector  space $S_{Dirac}$ of  
Dirac spinors. 
 
In particular,   the \blit{spin matrices}    $$\sigma _{\mu \nu}=\sigma _{[\mu \nu]} \equiv \frac{1}{2} [\gamma _{\mu},\gamma _{\nu}]_-  ,$$ defined for $\mu \ne \nu$,
  represent the bivectors, the Lie algebra generators  of the Spin group.    (An additional factor $i$ is generally introduced in quantum physics).  

The \blit{chirality   matrix} is defined as
\begin{equation}
\label{ }
\chi\equiv \gamma ^{5}\equiv \gamma _{5} \equiv - i~\gamma _{0}~\gamma _{1}~\gamma _{2}~\gamma _{3},\end{equation}
such that $\chi^2=\Id$. 

 {\bf  Explicit representations }

There are different ways to   represent the isomorphism $\Cliff(1,4) \mapsto \Mat_4(\setC)$.  The most familiar one is obtained  
as the complex $4\times 4$ matrices 
 \begin{equation}\label{DiracMatrices}
\gamma _{0}= 
{\left| 
\begin{array}{cc}\sigma _0 &0\\0&-\sigma _0 \end{array} \right|} ,
\gamma _{i}=
{\left| 
\begin{array}{cc}0&\sigma _i\\ -\sigma _i &0
\end{array} \right|}
,\end{equation}
where the $\sigma _{i}$ are the Pauli matrices. Their   products provide the rest of the basis  with, \eg, 
\begin{equation}
\label{ }
\gamma _{5}=i ~\gamma _{0}~\gamma _{1}~\gamma _{2}~\gamma _{3}={\left| 
\begin{array}{cc}0& \sigma _0 \\  \sigma _0   &0
\end{array} \right|}.
\end{equation}

Variants are found where $\gamma _{0}$ is replaced by $-\gamma _{0}$, or by $\matrixdd{0}{\Id}{\Id}{0}$; and where the $\gamma _{i}$ are replaced by the $-\gamma _{i}$.

Among other possibilities \cite{Rausch}, one may also   define $\gamma _{0}= 
{\left| 
\begin{array}{cc}0  &\sigma _0\\-\sigma _0&  0 \end{array} \right|} ,
$ with $  \gamma _{5}={\left| 
\begin{array}{cc}\sigma _0 &0\\0&-\sigma _0 \end{array} \right|} $.

\subsubsection{\kg ~and Dirac equations}

The \blit{\kg ~differential operator} $\Box + m^{2}$   is a second order  Lorentz-invariant operator. 
Historically, the spinors were introduced after   the quest  for  a first  order  Lorentz-invariant differential operator. 
This is only possible if the coefficients belong to a non commutative algebraic structure, which will be precisely (modulo isomorphisms) that of the \spt ~Clifford algebra. 
The new operator  acts on the spinor space both according to the spinor representation (since the spinor space is  a representation space for the \spt ~algebra) and differentially. The last action is to be understood as acting on spinor fields, \ie, \spt ~functions which take their values in $S$ (equivalently, sections of the \emph{spinor bundle}).

  The \kg ~operator  is factorized as  
the  product \begin{equation}
\Box + m^{2}=-(i\gamma ^{\mu} \partial _{\mu}+m)~(i   \gamma ^{\nu}  
\partial _{\nu}-m), \end{equation} where the four constant
$\gamma ^{\mu}$ 
verify  the conditions 
$$\gamma ^{\mu} \gamma ^{\nu} + \gamma ^{\nu} \gamma 
^{\mu}   = 2 \eta ^{\mu\nu}   \andy \partial _\mu \gamma ^{\nu}=0 $$  
(indices are lowered or raised with the \mink ~metric).

The usual field theory considers the $\gamma ^{\mu}$  as the  four matrices \blit{Dirac matrices}.  They correspond to  the first  grade  members of the \spt ~algebra. 


 \subsubsection{Projectors and  Weyl spinors}

The  complex $4\times 4$ matrices  matrices of this  representation   act on
 the space of  Dirac spinors $S_{Dirac} \simeq \setC ^{4}$. We write a Dirac spinor as $\psi_D$.
The $ \gamma _{\mu \nu} $ are the generators of  the   Lorentz rotations.

From the chirality matrix, one construct the two \blit{projectors}
 $$P_{left} \equiv \undemi~(\Id - \chi),~ P_{right} \equiv \undemi~(\Id +\chi)  .$$ 

They  project   a Dirac spinor $\psi_D$ onto its  right and left components $$
 \psi_R \equiv  P_{right}~  \psi_D=\vectord{0}{\eta },
 ~Ê\psi_L \equiv  P_{left}  ~ \psi_D=\vectord{\pi}{0};
 ~Ê \psi = \psi_R+ \psi_L =\vectord{\pi}{\eta}.$$ 
Thus,   $S_{Dirac}$  decomposes as 
$$S_{Dirac} =  P_+~  S_{Dirac}\oplus P_- ~Ê S_{Dirac},$$ 
where  
$ P_\pm  S_{Dirac}$ is the eigenspace of $\chi$ (in $S_{Dirac}$) with eigenvalue $\pm 1$ of the chirality operator. Elements of  
$ P_\pm  S_{Dirac}$ are called  right  (resp. left) helicity Weyl spinors, and they identify with the definitions above. Each 
$ P_\pm  S_{Dirac}$  is an irreducible  representation space for $\cclif ^{even}$ \cite{Castro}.
 
The Dirac and Weyl spinors also provide representations for the groups Pin and Spin.  

\subsection{Spinors in the  the \spt -algebra}

Spinors can be described as elements
of the \spt -algebra itself. Namely the [Dirac] spinor space identifies with a [left or right]Ê minimal ideal.
of the \spt algebra (\cite{Pavsic}  and references therein): the  [spin]Ê representation space
lies inside  the algebra   operating on it. 
 
Following \cite{Pavsic}, one may \eg ~select  the nilpotent element $f \eqbydef \unquart (e_0-e_3)~(e_1-i~e_2)$, and the vector space $C~ f$ appears as a minimal left ideal. This   4-dimensional vector space admits the  basis $(1, \theta_1 \eqbydef \undemi (e_0+e_3),   \theta_2 \eqbydef \undemi (e_1+i~Êe_2),   \theta_1  ~\theta_2)$.
It may be identified to the vector space $S_D$ of Dirac spinors. 
Then, $\gamma_5$ acts as the projector and generates the splitting $S_D=S_{left}\oplus S_{right}$  such that
$$i~\gamma_5~\psi _L=\psi _L;~~~~~
 i~\gamma_5~\psi _R=-\psi _R.$$

$S_{left}$, with  the basis $(1,  \theta_1  ~\theta_2)$, and $S_{right}$, with  the basis $(   \theta_1  ,\theta_2)$, may be seen as the Weyl spinor spaces in the \spt ~algebra.

Note that there are many different ways to embed the representation space into the \spt -algebra.

\section{The Clifford bundle on a [pseudo-]Riemanian manifold}

We have examined the \spt ~algebra, with its relations to spinors. Both were constructed over the \Minkvs, whose    affine version is the  the \minks.  Physical fields require the construction of \fb s over \spt, \ie, a pseudo-Riemannian manifold whose tangent spaces are copies of  the \Minkvs : the fibers are isomorphic to  the   Clifford algebra,  to the spinor spaces, to  the Spin groups ...  The basis is  \spt.  

A \spt ~is considered as four-dimensional orientable pseudo-Riemannian manifold. A  choice of time orientation (a polarization) allows to select those timelike or null vectors which are future directed.

\subsection{Fiber bundles associated to  a manifold}

To a differential  manifold $\calM$ are     associated \blit{natural vector  bundles} (\ie, on which the diffeomorphism group acts canonically),    among which the     tangent and the  cotangent bundles:
$$\T \calM \equiv \bigcup _{m\in \calM} \T _m \calM,  
~~Ê \T^* \calM \equiv \bigcup _{m\in \calM} \T^* _m \calM.$$
Each $\T _m \calM$ is a copy of the \Minkvs ~$\bbbM$, each 
  $\T^* _m \calM$ is a copy of its dual $\bbbM ^*$.

One defines the  $\binomial{n}{k}$ dimensional spaces of $k$-forms
 at $m$,  $ \bigwedge ^k (\T_m^* \calM)   $.  Their  union form the vector bundle of   $k$-forms on $\calM$, 
 $ \bigwedge ^k (\T ^*  \calM)   $.  
Its sections, the  element of  $\Sect(\bigwedge ^k \T_x^*  \calM)   $,  are the $k$-forms [fields] on $\calM$.

  At each point, one may define the two  Clifford algebras
  $ \Cliff(\T_m  \calM)$ and $ \Cliff(\T^*_m  \calM)$. Their  unions  define\begin{itemize}
  \item 
 the \blit{Clifford bundle of multivector fields}  on $\calM$: \\
$$\Cliff(\T   \calM) \equiv  \bigcup_m~ \Cliff(\T_m  \calM).$$ Each fibre     $\Cliff(\T_m \calM)$    is a copy
of the \spt ~algebra $\Cliff(1,3)$. 
  \item
   the \blit{Clifford bundle of    [differential]
multiforms} on $\calM$:  
   $$\Cliff(\T^*  \calM) \equiv  \bigcup_m~ \Cliff(\T_m^*  \calM).$$ Each  fibre    $\Cliff(\T_m^*  \calM)$   is also  a copy of $\Cliff(1,3)$.  $\Cliff(\T^*  M)$ is   similar to   the Cartan bundle $\bigwedge \T^*    \calM  \eqbydef   \cup_m ~\bigwedge 
 \T_m^*  \calM   $, the  difference lying  in the possibility of addition of   forms of different degrees  in $\Cliff(\T^*    \calM)$, not in 
$\bigwedge  \T^*    \calM$. The latter can be seen as embedded in  
$\Cliff(\T^*    \calM)$. 
 Here, $\cup_m ~\bigwedge 
 \T_m^*  \calM = \bigcup _{k=0}^n  \bigwedge ^k \T_x^*  M,$
where $\Sect(\bigwedge ^k \T_x^*  M)   $ is the $\binomial{n}{k}$ - dimensional space of $k$-forms.
\end{itemize}

 There is a complete isomorphism between 
 $\Cliff(\T^*  M)$ and $\Cliff(\T  M)$, which may be explicited by an extension of the canonical metric (musical) isomorphism.
 
 We recall that a scalar product in a vector space is naturally extended to the Clifford algebra (see \ref{Hodge}). Here, the metric of $\calM$ is extended 
to the tensor bundles, and thus to the  [sections of] the Clifford bundles, \ie, to multiforms and multivectors.

A metric compatible connection acts  on the tensor bundle. It is extended to  define  a covariant derivative acting on Clifford
fields (\ie, section of the Clifford bundles).

The Hodge duality, that we defined for arbitrary polyvectors (\ref{Hodge}) extends naturally to the Clifford bundles.

\subsection{Spin structure and spin bundle}
 
 We recall the principal  \fb s defined on a [pseudo-]Riemanian manifold:\begin{itemize}
 \item the \blit{frame bundle} $\Fr(\calM)$ (also written $\Fr\fibre \calM $)   is a GL-principal bundle on $\calM$, with structure group the general linear group GL. A  section is a   moving frame  of $\calM$:  a choice of a vector basis for the tangent space, at each point of the manifold.     
 \item the \blit{special orthogonal frame bundle} (or \blit{tetrad bundle})   $\FrSO \fibre \calM $ has structure group SO. 
A  section is an  oriented ON  frame  of $\calM$, or oriented  tetrad.   Orientability requires the vanishing  of  the first Stiefel--Whitney class.
  \item the \blit{time-oriented special orthogonal frame bundle} (or time-oriented\blit{tetrad bundle})    $\FrSOf     \fibre \calM $ has structure group $\SOf$.  A  section is an  oriented and time-oriented  ON  frame  of $\calM$.  
\end{itemize}
These bundles are well defined in a [pseudo-]Riemannian manifold. In a differential manifold, one may consider $\FrSO$  as the result of a process of \blit{\fb ~reduction} (see, \eg, \cite{Sardanashvily}) from $     \FrGL$,   equivalent to a choice of  metric.
 
 A \blit{Spin-structure} will be defined as a   Spin-principal \fb ~$\Fr^{\Spin} \fibre \calM $  called the \blit{Spin bundle} (also written  $\Spin    (\calM)$).
A section is called a \blit{spin frame}.
The Spin bundle is an  extension of $\FrSO \fibre \calM $ by the group $\setZ^2$. 
 
 All these $G$-principal \fb s have associated vector bundles with an action of the principal group $G$. The fibers are copies of a representation vector space of $G$. This   is the tangent bundle $\T \calM$ for the three first.  For $\Spin    (\calM)$,  they are  called the  \blit{spinor bundles}  (Weyl  spinor bundles, with  a  fibre  of complex dimension   2;    Dirac spinor bundles, with  a  fibre  of complex dimension 4). The fiber    
is a representation space for the group   Spin, \ie, a spinor space (see, \eg, \cite{Sharipov}).

{\bf Spin structure}

The Spin group is a the double \uc ~of the group SO. We recall  the double covering group  homomorphisms 
$$ H: \Spin \mapsto \SO,$$
$$H: \Spinf \mapsto  \SOf.$$

Given  a [pseudo-]Riemannian  manifold $\calM$, the special-orthogonal  bundle $\FrSO (\calM)$ is  a SO-principal \fb. A section is an oriented  ON frame (an oriented tetrad).
  
Even when $\Fr^{\SO} ({\calM})$ does exist, they may be some \blit{topological obstruction} to the existence of a spin structure. This  requires   the vanishing of the    second stiefel-Whitney class. 
Also, the existence of   a  spin structure is equivalent to the requirement that $\FrSO\fibre \calM $ is a trivial bundle. Since this is a principal bundle,  this means   that it  admits global sections (see, \eg, \cite{Rodrigues}), which are   global SO-tetrads. 
This implies that its \uc ~Ê$\Spin(\calM)$  also admits  global sections,  which are the \blit{spin frames}.

The   transition functions of the bundle  $f_{ij}$ of   $\FrSO \fibre \calM$   take their values in SO.  A spin structure  $\Spin(\calM)$, when it  exists,  is 
defined by its transition functions $\tilde{f}_{ij}$, with values in $\Spin $, and such that $H (\widetilde{f _{ij}})=f_{ij}$.
 Note that $\calM$   admits in general  many spin structures, depending on the choice of the 
$\widetilde{f _{ij}}$. 

The   2-1 homomorphism 
$$\widetilde{H}: ~Ê\Spin   (\calM) \mapsto \FrSO (\calM) $$   maps a fiber onto a fiber, so  that  $\widetilde{ H} (u~\lambda)=\widetilde{H} (u)~ H 
(\lambda),$ with $u \in \Spin   (\calM)$ (the \fb)
and $ \lambda \in \Spin   $ (the group). Considering the 4 connected 
components of SO,   the possible combinations  give  the 
possibility to 8 different spin structures corresponding to the 
choices of signs for $P^{2}$, $T^{2}$ and $(PT)^{2}$.
  
{\bf Proper spin structure}

The bundle 
$\FrSO (\calM)$ has for sections the oriented tetrads (= SO-frames). The bundle $ \FrSOf ({\calM})$  of {\sl time oriented} and oriented tetrads  is obtained after  the [\fb]  reduction  of the Lorentz group  SO  to the  {\sl proper} Lorentz 
 group $\SOf$. Then a \blit{proper spin structure}
is defined as a principal bundle  $\Spinf (\calM)$ over $\calM$, with structure group 
$\Spinf$. 

{\bf Spinor fields}

The group Spin
acts on its representations which are the spinor [vector]Êspaces.   A vector bundle over \spt,  whose fibre is such a representation  space for  the group  Spin is a   \blit{spinor bundle}; a section is  called a   \blit{spinor field}. This is a spinor-valued function (0-form) on \spt.  
In particular, the Dirac and the  (left and right) Weyl  spinor bundles have for sections 
the corresponding  Dirac  or Weyl \emph{spinor fields}. Representations of the  group Spin act on them.

  {\bf Spin connections}

A  \blit{linear connection} on a differential manifold $\calM$ identifies with a principal connection on the principal \fb ~$\Fr \fibre \calM$, with the linear group  GL as principal group.
A metric structure allows (is equivalent to) a \fb ~bundle reduction 
from $\Fr \fibre \calM$ to the orthogonal frame bundle
$\FrSO \fibre \calM$. 
The linear connection (with values in the Lie algebra $\Liegl$) is reduced to a  \blit{Lorentz connection}, with values in the Lie algebra $\Lieso$. 
This   is a principal connection   on the principal \fb ~$\FrSO \fibre \calM$.
When a spin structure exists, the latter defines a connection 
of the principal \fb ~$\Fr^{\Spin} \fibre \calM$, which is called a \emph{spin connection}.

{} 
 
\end{document}